\documentclass{ELSP}
\PassOptionsToPackage{unicode}{hyperref}
\PassOptionsToPackage{hyphens}{url}
\IfFileExists{upquote.sty}{\usepackage{upquote}}{}
\IfFileExists{microtype.sty}{% use microtype if available
	\usepackage[]{microtype}
	\UseMicrotypeSet[protrusion]{basicmath} % disable protrusion for tt fonts
}{}
\makeatletter
\IfFileExists{parskip.sty}{%
	\usepackage{parskip}
}
\makeatother
\IfFileExists{xurl.sty}{\usepackage{xurl}}{} % add URL line breaks if available
\IfFileExists{bookmark.sty}{\usepackage{bookmark}}{\usepackage{hyperref}}
\hypersetup{
	pdfcreator={ELSP}}
\usepackage{longtable,booktabs,array}
\usepackage{calc} % for calculating minipage widths
\usepackage{etoolbox}
\makeatletter
\patchcmd\longtable{\par}{\if@noskipsec\mbox{}\fi\par}{}{}
\makeatother
\IfFileExists{footnotehyper.sty}{\usepackage{footnotehyper}}{\usepackage{footnote}}
\makesavenoteenv{longtable}
\usepackage{setspace}
\usepackage{graphicx}
\usepackage{amsmath}
\usepackage{times}
\usepackage{mathptmx}

\DeclareMathAlphabet{\mathcal}{OMS}{cmsy}{m}{n}
\DeclareSymbolFont{largesymbols}{OMX}{cmex}{m}{n}

\usepackage{color}
\usepackage{caption}
\usepackage{subcaption}
\usepackage{geometry}
\usepackage{float}
\usepackage{enumerate}
\usepackage{enumitem}
\usepackage{stfloats}
\usepackage{ragged2e}
\usepackage{titlesec}
\usepackage{ragged2e}
\usepackage{amsmath,amsfonts}
\usepackage{booktabs}
\usepackage{multirow}
\usepackage{graphicx}
\usepackage{enumitem}
\usepackage{subcaption}
\usepackage{xspace}
\usepackage{tikz}
\usetikzlibrary{shapes,arrows,positioning,fit,calc,decorations.pathreplacing}
\usepackage{pgfplots}
\usepgfplotslibrary{polar}
\pgfplotsset{compat=1.17}
\usepackage{hyperref}
\usepackage{doi}
\usepackage{cleveref}
\usepackage{balance}

\titleformat{\section}[hang]
  {\fontsize{12pt}{12pt}\selectfont\bfseries\color[RGB]{0,131,255}} 
  {\thesection}{0.5em}{}

\titleformat{\subsection}[hang]
  {\fontsize{12pt}{12pt}\selectfont\itshape} 
  {\thesubsection}{0.5em}{}

\titleformat{\subsubsection}[hang]
  {\fontsize{12pt}{12pt}\selectfont} 
  {\thesubsubsection}{0.5em}{}

\usepackage{fancyhdr}
\fancypagestyle{firstpage}
{
	\fancyhf{}
	\setlength{\headsep}{6px}
	\fancyhead[R]{\sffamily \fontsize{9pt}{8pt}\selectfont \textbf{\textcolor[RGB]{0,131,255}{Journal}}}
	\fancyhead[L]{\sffamily \fontsize{9pt}{8pt}\selectfont \textbf{\textcolor[RGB]{0,131,255}{ELSP}}}
	\fancyfoot[L]{\sffamily \fontsize{9pt}{8pt}\selectfont Author {\em et al}. Journal Year (Issue): XXXX}
}

\hypersetup{
    colorlinks=false, % 
    pdfborder={0 0 0}, % 
    hidelinks % 
}
\usepackage[utf8]{inputenc}
\usepackage[table,xcdraw,RGB]{xcolor}
\usepackage{enumitem}
\usepackage[numbers,sort,compress]{natbib} 
\usepackage{xspace}
\usepackage{natbib}
\setcitestyle{numbers,square,comma,sort&compress}
\newlength\mylen
\newlength\mybiblabelwidth
\setcitestyle{citesep={,\kern-\mylen}}

\begin{document}

%% ===== Custom Commands =====
\newcommand{\ie}{\textit{i.e.}\xspace}
\newcommand{\eg}{\textit{e.g.}\xspace}
\newcommand{\etal}{\textit{et al.}\xspace}
\newcommand{\abim}{\textsc{ABIM}\xspace}
\newcommand{\dirA}{\textbf{B\,$\boldsymbol{\rightarrow}$\,A}\xspace}
\newcommand{\dirB}{\textbf{A\,$\boldsymbol{\rightarrow}$\,B}\xspace}

%\hfill \textbf{\large Type of the paper}
\thispagestyle{firstpage}

\let\thefootnote\relax
\footnotetext{
\newline
\vspace{12pt}
    \raisebox{-\height}[0pt][0pt]{ % 精确控制垂直位置
        \begin{minipage}[h]{\linewidth}
            \begin{minipage}[h]{0.15\linewidth}
                \includegraphics[width=\linewidth, height=1cm]{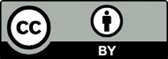} % 控制图片尺寸
            \end{minipage}
            \hfill
            \begin{minipage}[h]{0.82\linewidth}
            \justifying
                \footnotesize Copyright©Year by the authors. Published by ELSP. 
                This work is licensed under a Creative Commons Attribution 4.0 
                International License, which permits unrestricted use, distribution, 
                and reproduction in any medium provided the original work is properly cited.
            \end{minipage}
        \end{minipage}
    }
}
 % 减少页面可用高度，将内容上推
 % 确保从顶部开始

\setstretch{1.24}
\begin{flushleft}
% \papertype{paper type}
{\sffamily \small \noindent {\textls[-40]{Paper Type $\mid$ Received Day Mon Year; Revised Day Mon Year; Accepted Day Mon Year; Published Day Mon Year}}}\\[-0.7ex]
%Example
%\timeline{15 May 2022}{17 July 2022}{20 September 2022}
{\sffamily\small{https://doi.org/10.55092/xxxx}}

\vspace{12pt}
{\raggedright
\papertitle{Toward Web 4.0: Bidirectional Trust between AI Agents and Blockchain%
\hfill
\href{https://crossmark.crossref.org/dialog/?doi=10.55092/XXXX}{%
\raisebox{-0.2em}{\includegraphics[width=0.93cm,height=0.93cm]{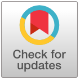}}%
}}
}
 
% \newcontent The title should be concise, informative and meaningful. It should
% include key terms to make it easier to be found via searching. Please
% avoid long systemic names, obscure abbreviations, acronyms or symbols or
% formulas. Avoid phrases such as ``on the'', ``a study of'', ``research
% on'', ``report on'' ``regarding'', and ``use of'', omit ``the'' at the
% beginning of the title.

\vspace{12pt}
\authorname{Yunfeng} {Xia} {1},
\authornameCorres{Chao} {Li}{1,}{*}
\authorname{Lei} {Li} {1}
\authorname{Chenhao} {Zhang} {1}
\authorname{Li} {Duan} {1}
\authorname{Runhua} {Xu} {2}
\textbf{and} 
\authorname{Wei} {Wang} {3}

\vspace{12pt}
\formatintroduction{1}{Beijing Key Laboratory of Security and Privacy in Intelligent Transportation, Beijing
Jiaotong University, Beijing, China}
\formatintroduction{2}{Beihang University, Beijing, China}
\formatintroduction{3}{Xi'an Jiaotong University, Xi'an, China}
% \formatintroduction{2}{Department, Institution, City, Country}

% \newcontent Note: Pease list all authors' full names and institutions. If an
% author's current address is different from the address where the work
% was carried out, please add note. The general note symbol should be used
% in the following order: *,†,‡,§,¶,**,††,‡‡. Author Contribution:
% we encourage authors to make specific attributions of contribution and
% responsibility in the acknowledgements of the article, otherwise all
% co-authors will be taken to share full responsibility for all of the
% paper. Authors may wish to use a taxonomy such as
% \href{http://credit.niso.org/}{CRediT} to describe the contributions of
% each author.

\vspace{12pt}
\authoremail{Correspondence author(s)} {li.chao@bjtu.edu.cn.}
\end{flushleft}

% \vspace{12pt}
% \noindent\textbf{\textcolor[RGB]{0,131,255}{Highlights:}}
% \vspace{12pt}
% \begin{itemize}[left=0pt, labelwidth=0pt, labelsep=17pt, itemsep=0pt]
%     \item Highlight 1
%     \item Highlight 2
%     \item Highlight 3
% \end{itemize}

\vspace{10pt}
\noindent\textbf{\textbf{\textcolor[RGB]{0,131,255}{Abstract:}}} Autonomous AI agents are increasingly deployed on blockchain platforms, yet the design space that governs their interaction remains poorly understood. This convergence, where autonomous agents operate on and within decentralized systems, is a defining feature of the emerging Web~4.0 paradigm. This paper presents a Systematization of Knowledge (SoK) organized around a \emph{bidirectional trust framework}. In the \dirA direction, we examine how blockchain provides trust infrastructure for agents, spanning identity and account abstraction, permission and delegation, intent-centric execution, and tokenized agent economies. In the \dirB direction, we examine the reverse: how AI agents participate in core blockchain mechanisms including security auditing, consensus, and governance. A Trust Foundation of verifiable computation (zkML, opML, TEE) underpins both directions, with each primitive offering different trade-offs between trust minimality, computational overhead, and deployment readiness. We formalize the interaction as an Agent-Blockchain Interaction Model (\abim), catalog 70 Ethereum EIPs/ERCs (Appendix~\ref{app:eip-catalog}), examine 20 representative industry projects, and review 118 academic papers, applying a five-dimensional framework assessing Verifiability, Minimality of Trust, Expressiveness, Composability, and Maturity. Our analysis uncovers significant gaps: the agent-specific standards ecosystem is overwhelmingly immature (only 2 of 13 direct AI/agent ERCs have reached Final), intent architectures lack formal analysis, and while isolated works have begun to explore AI participation in consensus and governance, a unified security framing that treats AI as a first-class actor at the protocol layer remains absent. We propose a three-dimensional taxonomy (agent autonomy $\times$ trust model $\times$ operational direction), identify nine concrete open problems, and highlight the sharpest research opportunities at this intersection.

\vspace{+12pt}
\noindent\textbf{\textcolor[RGB]{0,131,255}{Keywords:}} AI Agent, Blockchain, Web 4.0, Account Abstraction, Intent Architecture, Verifiable Computation, DAO Governance, Trust Framework, Systematization of Knowledge

%% ==========================================================
%% SECTION 1: INTRODUCTION
%% ==========================================================
\section{Introduction}\label{sec:intro}

Large language models and autonomous AI agent frameworks have reshaped how intelligent systems interact with digital infrastructure\cite{karim2025multi-agent}. At the same time, blockchain technology has matured beyond simple value transfer into a programmable trust layer\cite{werner2021sok-defi}: smart contracts on Ethereum-like chains now support sophisticated execution patterns such as privacy-preserving timed transactions~\cite{twatch}, Delegated Proof-of-Stake (DPoS) and related consensus designs power widely-used Web~3.0 networks including EOSIO, Steem, and TRON despite known governance-takeover risks~\cite{DPoS,steemb}, and an application stack above these layers now spans cross-chain access control with decentralized storage~\cite{iotj}, NFT marketplaces~\cite{wyfnft}, and Web3 social platforms~\cite{wbfritech}. Where these two trajectories converge, with autonomous AI agents that operate on and within blockchain systems, a rapidly growing design space is emerging, one with deep implications for trust, security, and economic coordination. This convergence is increasingly recognized as a defining characteristic of Web~4.0: the next evolution of the internet in which autonomous AI agents, rather than human users, become first-class network participants that interact, transact, and coordinate through decentralized infrastructure~\cite{zhou2025web4}.

The urgency of systematizing this space stems from three converging pressures. First, the \emph{economic stakes are already substantial}: as of December 2024, the aggregate market valuation of blockchain-based AI agents reached an estimated \$8.6 billion~\cite{ante2025agent-defi}, yet the interaction between AI-assisted development and security-critical smart contracts has already surfaced in high-profile incidents. For example, an oracle misconfiguration in the Moonwell protocol resulted in a \$1.78M loss; a pull request associated with the affected component carried a \texttt{Co-Authored-By} trailer indicating AI tooling involvement, though Moonwell's official post-mortem did not attribute the root cause to AI-generated code~\cite{moonwell2026}. Second, the \emph{trust gap is widening}: as agents transition from read-only analytics to autonomous transaction execution, the absence of formal trust models means that each project invents ad hoc security assumptions, creating a fragmented landscape where vulnerabilities compound across layers. Third, the \emph{academic vacuum is becoming costly}: with no prior SoK covering account abstraction, only one formal analysis of intent architectures~\cite{chitra2024intent-markets}, and no systematic framing for AI participation in consensus or governance, practitioners lack the theoretical foundations needed to reason about safety, incentive compatibility, and failure modes, precisely the properties that matter most as agent autonomy increases.

The scale of this convergence is already substantial. Ante~\cite{ante2025agent-defi} analyzes 306 projects that combine AI and blockchain technology. Ethereum alone hosts more than 70 EIPs and ERCs that directly or indirectly affect AI agent operations, covering account abstraction~\cite{erc4337,eip7702}, agent identity registration~\cite{erc8004,erc8126}, verifiable ML inference~\cite{erc7992,erc7007}, intent-centric execution~\cite{erc7521,erc7683}, smart contract delegation~\cite{erc7710}, and agent tokenization~\cite{erc7662,erc7857}. Despite this explosive industry growth, the academic landscape has not kept pace: no prior SoK covers account abstraction, intent-driven architectures have received only one formal analysis~\cite{chitra2024intent-markets}, and although isolated studies have examined AI in consensus~\cite{squirrl2021} and governance~\cite{agentdao2025}, a systematic security framing for AI agents as \emph{participants}, rather than mere tools, in blockchain protocol mechanisms remains absent.

Our central thesis is that the relationship between AI agents and blockchain is \emph{bidirectional and symbiotic}. In what we call the \dirA direction (Blockchain $\to$ Agent), blockchain provides trust infrastructure for agents, including identity, permissioned execution, intent specification, and economic incentives. In the reverse \dirB direction (Agent $\to$ Blockchain), agents bring intelligent capabilities to the blockchain itself through automated security auditing, consensus participation, and governance decision-making. Underpinning both directions is a foundation of \emph{verifiable computation}, the ability to cryptographically prove that an agent's reasoning and actions are correct without trusting any single party. The degree to which verifiable computation is needed depends on application context: high-stakes autonomous operations demand strong cryptographic guarantees, while lower-risk assistive agents may operate acceptably under weaker trust models such as reputation or economic bonds.

\begin{figure*}[t]
\centering
\begin{tikzpicture}[
    layer/.style={draw, rounded corners=3pt, minimum width=5.8cm, minimum height=1.05cm,
                  font=\small, align=center, thick},
    foundation/.style={draw, rounded corners=4pt, minimum width=13.6cm, minimum height=1.3cm,
                       fill=black!8, font=\small, align=center, thick},
    arrowlbl/.style={font=\footnotesize\itshape, text=black!70},
    dirtitle/.style={font=\small\bfseries, text=black!80},
]

%% --- Trust Foundation (bottom) ---
\node[foundation] (tf) at (0,0) {
    \textbf{Trust Foundation: Verifiable Computation}\\[-1pt]
    {\footnotesize zkML\;\textbullet\;opML\;\textbullet\;TEE\;\textbullet\;Fraud Proofs\;\textbullet\;Formal Verification}
};

%% --- Direction A (left column) ---
\node[layer, fill=blue!8] (a1) at (-3.6, 1.8) {\textbf{A1.} Account \& Identity\\[-2pt]{\footnotesize AA $\cdot$ Agent Registration $\cdot$ DID}};
\node[layer, fill=blue!12] (a2) at (-3.6, 3.1) {\textbf{A2.} Permission \& Delegation\\[-2pt]{\footnotesize Modular Accounts $\cdot$ Session Keys}};
\node[layer, fill=blue!16] (a3) at (-3.6, 4.4) {\textbf{A3.} Intent \& Execution\\[-2pt]{\footnotesize Intent Specs $\cdot$ Solvers $\cdot$ MEV}};
\node[layer, fill=blue!20] (a4) at (-3.6, 5.7) {\textbf{A4.} Agent Economy\\[-2pt]{\footnotesize Tokenization $\cdot$ Markets $\cdot$ IMO}};

%% --- Direction B (right column) ---
\node[layer, fill=orange!10] (b1) at (3.6, 1.8) {\textbf{B1.} AI + Security\\[-2pt]{\footnotesize Audit $\cdot$ Monitoring $\cdot$ Defense}};
\node[layer, fill=orange!15] (b2) at (3.6, 3.55) {\textbf{B2.} AI + Consensus\\[-2pt]{\footnotesize Validators $\cdot$ PoL $\cdot$ BFT}};
\node[layer, fill=orange!20] (b3) at (3.6, 5.3) {\textbf{B3.} AI + Governance\\[-2pt]{\footnotesize DAO Voting $\cdot$ Proposals $\cdot$ Legal}};

%% --- Direction labels ---
\node[dirtitle] at (-3.6, 6.6) {B\,$\to$\,A: Blockchain $\to$ Agent};
\node[dirtitle] at (3.6, 6.6) {A\,$\to$\,B: Agent $\to$ Blockchain};

%% --- Arrows from foundation ---
\draw[->, thick, blue!50] (tf.north -| a1.south) -- (a1.south);
\draw[->, thick, orange!60] (tf.north -| b1.south) -- (b1.south);

%% --- Bidirectional trust arrows between A and B (horizontal) ---
\foreach \y in {2.5, 3.75, 5.0} {
    \draw[<->, thick, black!50, dashed] (-0.7, \y) -- (0.7, \y);
}
% \node[font=\scriptsize\itshape, text=black!60] at (0, 7.0) {$\longleftrightarrow$ Bidirectional Trust};
\node[font=\scriptsize\itshape, text=black!60] at (0, 6.4) {Bidirectional};
\node[font=\scriptsize\itshape, text=black!60] at (0, 6.1) {Trust};
\node[font=\scriptsize\itshape, text=black!60] at (0, 5.8) {$\longleftrightarrow$};

%% --- Verification links (dotted) ---
\draw[dotted, thick, black!30] (tf.north west) ++(0.3,0) -- ++(0,6.2);
\draw[dotted, thick, black!30] (tf.north east) ++(-0.3,0) -- ++(0,6.2);

\end{tikzpicture}
\caption{Bidirectional Trust Framework for autonomous AI agents on blockchain. The \dirA direction (left): blockchain provides trust infrastructure for agents across four layers. The \dirB direction (right): agents participate in core blockchain mechanisms across three layers. The Trust Foundation of verifiable computation underpins both directions, with varying necessity depending on agent autonomy and application risk.}
\label{fig:framework}
\end{figure*}
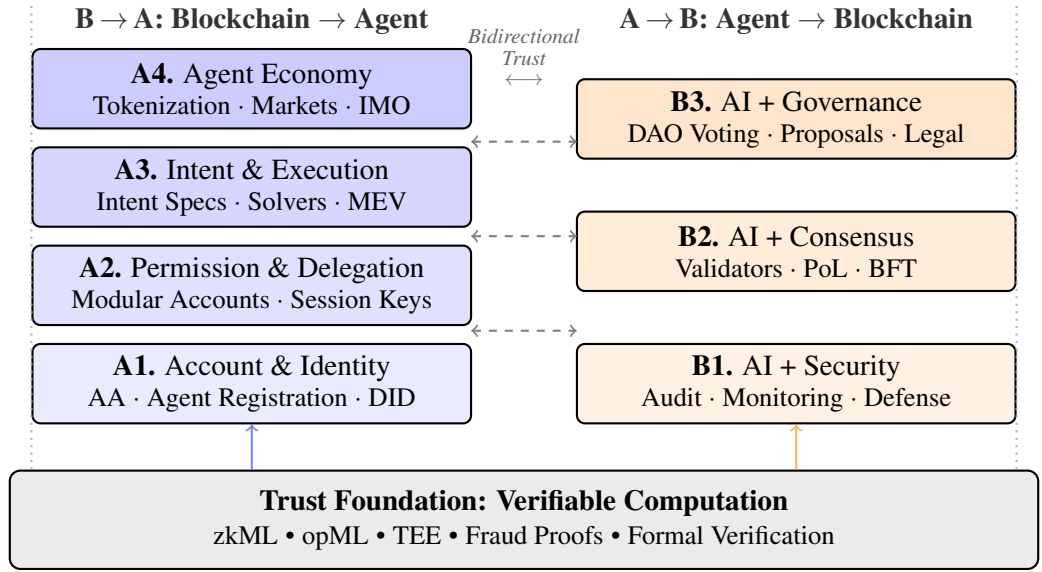

We organize the \dirA direction into four layers that mirror the lifecycle of an agent's on-chain operation, each building on the previous one: an agent must first \emph{exist} on-chain with a verifiable identity (\textbf{A1:~Account \& Identity}), then be \emph{authorized} to act within bounded permissions (\textbf{A2:~Permission \& Delegation}), then \emph{express} what it wants to achieve through intent specifications (\textbf{A3:~Intent \& Execution}), and finally \emph{be economically sustained} through tokenization and incentive mechanisms (\textbf{A4:~Agent Economy}). This ordering reflects a strict \emph{dependency chain}: A2 cannot function without A1 (permissions require identity), A3 presupposes A2 (intent execution requires authorization), and A4 builds on all three (economic participation requires identity, permissions, and execution capability). The \dirB direction is organized into three layers ordered by increasing depth of participation: agents first \emph{observe and audit} blockchain activity (\textbf{B1:~Security}), then potentially \emph{validate} transactions as consensus participants (\textbf{B2:~Consensus}), and ultimately \emph{shape} protocol evolution through governance (\textbf{B3:~Governance}). The \textbf{Trust Foundation~(TF)} of verifiable computation is not a separate layer but a \emph{cross-cutting substrate}: it provides the correctness guarantees that all other layers may draw upon, with the strength of guarantee varying by application risk and agent autonomy. \Cref{fig:framework} visualizes this architecture; the vertical arrows from TF to both directions indicate this foundational role, while the horizontal dashed arrows between corresponding A and B layers reflect that the two directions are mutually reinforcing rather than independent, since, for example, the same identity infrastructure (A1) that enables agent operations also enables verifiable participation in governance (B3).

This paper makes six contributions.
\emph{(i)}~We formalize the interaction through an Agent-Blockchain Interaction Model (\abim) that captures the trust assumptions and security properties at each layer (\S\ref{sec:background}).
\emph{(ii)}~We introduce a five-dimensional evaluation framework (Verifiability, Minimality of Trust, Expressiveness, Composability, and Maturity) for systematically comparing EIPs, industry projects, and academic proposals.
\emph{(iii)}~We provide a cross-cutting synthesis (\S\ref{sec:crosscutting}) that jointly analyzes 70 EIPs/ERCs, \abim compliance across all layers, and 20 representative industry projects, exposing ecosystem-level (mis)alignments invisible to single-layer analysis.
\emph{(iv)}~We move beyond the prevailing ``blockchain as tool for AI'' perspective to examine AI agents as active participants in consensus and governance, a dimension absent from prior surveys (\S\ref{sec:direction-b}).
\emph{(v)}~We develop a threat taxonomy covering seven attack categories and identify nine concrete open research problems (\S\ref{sec:open-problems}).
\emph{(vi)}~We construct a three-dimensional taxonomy (agent autonomy $\times$ trust model $\times$ operational direction) that maps the ecosystem and highlights under-explored opportunities (\S\ref{sec:taxonomy}).

To situate this work within the existing literature, \Cref{tab:differentiation} compares our coverage with four recent surveys. Ante~\cite{ante2025agent-defi} provides the most comprehensive project catalog (306 entries) and proposes a governance framework mapping autonomy against decentralization, but does not examine EIP/ERC standards, formal trust models, or the \dirB direction. Karim~et~al.~\cite{karim2025multi-agent} address multi-agent security and scalability but do not analyze EIP/ERC standards. Romandini~et~al.~\cite{sok-ai-blockchain-security2025} propose a four-layer software architecture and a three-class agent taxonomy, and offer the deepest treatment of security and privacy; however, their architecture describes how agent systems are built rather than what trust guarantees each layer provides, and they cover only the \dirA direction (agents using blockchain), omit account abstraction and intent architectures entirely, and do not analyze EIP/ERC standards or formalize security properties. The closest work~\cite{autonomous-agents-blockchains2026} provides the most detailed standards analysis with a five-part integration taxonomy and a comparative capability matrix of 85 systems across 13 dimensions, but does not cover the \dirB direction, does not systematize verifiable computation (zkML/opML/TEE), and does not propose a formal interaction model. Ours is, to our knowledge, the first survey to systematize both directions, evaluate all agent-relevant EIPs/ERCs through a unified five-dimensional framework, and formalize the interaction as \abim.

\begin{table}[t]
\caption{Comparison with related surveys. $\bullet$ = comprehensive, $\circ$ = partial, --- = not covered.}
\label{tab:differentiation}
\centering
\begin{tabular}{lccccc}
\toprule
Dimension & \rotatebox{70}{\cite{ante2025agent-defi}} & \rotatebox{70}{\cite{karim2025multi-agent}} & \rotatebox{70}{\cite{sok-ai-blockchain-security2025}} & \rotatebox{70}{\cite{autonomous-agents-blockchains2026}} & \rotatebox{70}{\textbf{Ours}} \\
\midrule
Account\ Abstraction (A1) & --- & --- & $\circ$ & $\bullet$ & $\bullet$ \\
Permission (A2) & --- & --- & --- & $\circ$ & $\bullet$ \\
Intent Architecture (A3) & --- & --- & --- & $\circ$ & $\bullet$ \\
Agent Economy (A4) & $\bullet$ & --- & --- & $\circ$ & $\bullet$ \\
AI + Security (B1) & --- & $\circ$ & $\bullet$ & --- & $\bullet$ \\
AI + Consensus (B2) & --- & $\circ$ & --- & --- & $\bullet$ \\
AI + Governance (B3) & --- & --- & --- & --- & $\bullet$ \\
Verifiable Computation (TF) & --- & --- & $\circ$ & $\circ$ & $\bullet$ \\
Formal Model & --- & --- & --- & --- & $\bullet$ \\
EIP/ERC Analysis & --- & --- & --- & $\bullet$ & $\bullet$ \\
Taxonomy & $\circ$ & $\circ$ & $\circ$ & $\bullet$ & $\bullet$ \\
Threat Model & --- & $\circ$ & $\bullet$ & $\bullet$ & $\bullet$ \\
\bottomrule
\end{tabular}
\end{table}

The remainder of the paper is organized as follows. \S\ref{sec:background} provides background and formal definitions. \S\ref{sec:trust-foundation} examines the Trust Foundation layer of verifiable computation. \S\ref{sec:direction-a} and \S\ref{sec:direction-b} cover the \dirA and \dirB directions, respectively. Having traversed the framework layer by layer, \S\ref{sec:crosscutting} then shifts to a cross-cutting view that synthesizes findings across all layers, analyzing privacy and cross-chain tensions, standards maturity and dependency structure, \abim security-property compliance, and the industry project landscape, so as to expose ecosystem-level patterns that single-layer analysis cannot reveal. Building on this synthesis, \S\ref{sec:taxonomy} develops the three-dimensional taxonomy, and \S\ref{sec:open-problems} identifies open problems. \Cref{fig:framework} illustrates the overall architecture.

%% ===== Fig 1: Bidirectional Trust Framework =====

%% ==========================================================
%% SECTION 2: BACKGROUND AND PROBLEM DEFINITION
%% ==========================================================
\section{Background and Problem Definition}\label{sec:background}

\subsection{From EOA to AA to Agent: Three Generations of Blockchain Accounts}\label{sec:bg-accounts}

Understanding how AI agents interact with on-chain systems requires tracing the evolution of blockchain account models through three distinct generations.

Since Ethereum's inception in 2015, the dominant account type has been the Externally Owned Account (EOA), controlled by a single secp256k1 private key. EOAs impose rigid constraints: they support only one signature scheme, cannot implement custom validation logic, and require the key holder to individually sign every transaction. These limitations make automated or delegated execution impossible at the protocol level, a serious obstacle for any agent that needs to act on a user's behalf. The concept of account abstraction (AA) relaxes these constraints by allowing smart contracts to serve as first-class accounts. Throughout this paper we annotate each standard with its EIP process status~\cite{eip1}: \emph{Draft}, \emph{Review}, \emph{Last Call}, \emph{Final} (stable specification), \emph{Stagnant} (inactive $\geq$6 months), or \emph{Withdrawn}; note that specification maturity does not imply adoption, as many Draft-stage standards are already deployed in production. The idea dates back to EIP-86~\cite{eip86} (2017, now Stagnant), which proposed abstracting transaction origin and signature verification, and was revisited in EIP-2938~\cite{eip2938} (2020, Withdrawn), which aimed at native consensus-layer support. The current dominant standard, ERC-4337~\cite{erc4337} (2021, Review), takes a different path: it introduces an alternative mempool of \texttt{UserOperations} that are batched by \texttt{Bundlers} and executed through a singleton \texttt{EntryPoint} contract, all without requiring consensus-layer changes. Wang and Chen~\cite{wang2023aa-analysed} provide a structured analysis of this architecture, while measurement studies~\cite{lin2024erc4337-measurement} show that creating a \texttt{SimpleAccount} costs 381,489 gas, roughly 18 times a standard EOA transfer (21{,}000 gas). More recently, EIP-7702~\cite{eip7702} (2024, Final), introduced in the Pectra upgrade, allows EOAs to set contract code, blurring the boundary between the two account types. However, security analysis by Qi~\etal~\cite{eip7702-phishing2025} reveals that the interaction between EIP-7702 and ERC-4337 creates compound attack vectors not present in either standard alone.

\begin{figure}[t]
\centering
\begin{tikzpicture}[
    x=0.76cm,
    milestone/.style={circle, draw, fill=white, inner sep=0pt, minimum size=6pt, thick},
    gen/.style={rounded corners=3pt, draw, thick, minimum height=0.7cm, align=center, font=\scriptsize},
    yr/.style={font=\tiny, text=black!60},
    lbl/.style={font=\scriptsize, align=center},
]

%% --- Main axis ---
\draw[thick, ->, black!50] (0,0) -- (10.5,0);

%% --- Year ticks ---
\foreach \x/\y in {0.5/2015, 2/2017, 3.5/2020, 4.8/2021, 7/2024, 9/2025, 10/2026} {
    \draw[black!40] (\x, -0.1) -- (\x, 0.1);
    \node[yr, below] at (\x, -0.15) {\y};
}

%% --- Generation bands ---
\node[gen, fill=blue!6, minimum width=1.0cm] at (0.5, 1.6) {\textbf{Gen\,1}: EOA};
\node[gen, fill=blue!12, minimum width=3.3cm] at (4.4, 1.6) {\textbf{Gen\,2}: Account Abstraction};
\node[gen, fill=orange!12, minimum width=2.0cm] at (9, 1.6) {\textbf{Gen\,3}: Agent};

%% --- Milestones ---
\node[milestone, fill=blue!20] (e86) at (2, 0) {};
\node[lbl, above=3pt] at (e86) {EIP-86};

\node[milestone, fill=blue!20] (e2938) at (3.5, 0) {};
\node[lbl, below=12pt] at (e2938) {EIP-2938};

\node[milestone, fill=blue!40] (e4337) at (4.8, 0) {};
\node[lbl, above=3pt] at (e4337) {ERC-4337};

\node[milestone, fill=blue!50] (e7702) at (7, 0) {};
\node[lbl, above=3pt] at (e7702) {EIP-7702};

\node[milestone, fill=orange!40] (e8004) at (8.2, 0) {};
\node[lbl, below=12pt] at (e8004) {ERC-8004};

\node[milestone, fill=orange!40] (e8126) at (9.2, 0) {};
\node[lbl, above=3pt] at (e8126) {ERC-8126};

\node[milestone, fill=orange!40] (e7857) at (10, 0) {};
\node[lbl, below=12pt] at (e7857) {ERC-7857};

%% --- Evolution arrows ---
\draw[->, thick, black!25, dashed] (e86) to[bend left=20] (e2938);
\draw[->, thick, black!25, dashed] (e2938) to[bend left=20] (e4337);
\draw[->, thick, black!30, dashed] (e4337) to[bend left=15] (e7702);

\end{tikzpicture}
\caption{Evolution of blockchain account models: EOA (Gen 1), Account Abstraction proposals (Gen 2), and agent-specific identity standards (Gen 3).}
\label{fig:timeline}
\end{figure}
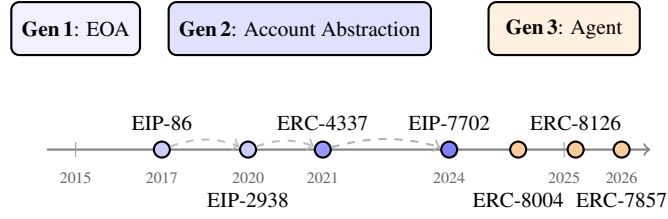

The latest wave of standards moves beyond general account abstraction to address AI agent identity specifically. ERC-8004~\cite{erc8004} defines an agent discovery protocol with identity, reputation, and validation registries. ERC-8126~\cite{erc8126} enables agents to self-register with multi-layer verification and ZKP-based privacy, including a quantitative risk score ranging from 0 to 100. ERC-7857~\cite{erc7857} takes a different approach, representing agents as NFTs whose model parameters and metadata can be encrypted using TEE and ZKP. The majority of these third-generation standards have not yet reached Final status, a point we revisit in \S\ref{sec:maturity}. Among the projects we surveyed, PAN Network~\cite{pan2025} is one of the few that explicitly implements ERC-8004, combining it with an HTTP-native x402 payment protocol to enable agent-to-agent service discovery, price negotiation, and settlement. DeAgentAI~\cite{deagentai2024} addresses the same triad of challenges, namely identity, consensus for probabilistic outputs, and state continuity, through sovereign identity on Sui and BSC. Additional standards address agent consent and interoperability: ERC-7517~\cite{erc7517} defines metadata fields for declaring whether an agent's outputs may be used for AI/ML training, ERC-8033~\cite{erc8033} introduces a multi-agent council oracle model with commit-reveal and staking incentives, and ERC-7913~\cite{erc7913} enables signature verification for non-secp256k1 keys (including TEE-generated keys), broadening the authentication mechanisms available to agents. Recent academic work further explores unified identity delegation frameworks~\cite{agent-identity-delegation2026}, DID/VC-based authentication~\cite{agent-did-vc2025}, and the intersection of proof-of-personhood with ZKP for AI alignment~\cite{pop-zkp-alignment2025}. \Cref{fig:timeline} visualizes this three-generation evolution.

%% ===== Fig 2: Account Evolution Timeline =====

\subsection{From Rule-Based to LLM-Agent: The Evolution of AI Autonomy}\label{sec:bg-ai}

The evolution of AI techniques applied to blockchain can be broadly characterized as a progression through three overlapping phases, a pattern most clearly documented in the smart contract security literature~\cite{debaets2024vulnerability,blockscan2024}. Early on-chain automation relied on deterministic rule-based scripts, which were fully verifiable but extremely limited in scope. Subsequently, ML-assisted systems introduced learned models for tasks such as vulnerability detection and anomaly identification, though humans remained in the execution loop. Most recently, LLM-powered autonomous agents have begun to independently interpret, plan, and execute multi-step operations, handling everything from DeFi trading to governance participation~\cite{karim2025multi-agent}.

A crucial observation runs through this progression: each generation increases \emph{expressiveness} while \emph{decreasing verifiability}. A rule-based bot can be formally verified against its specification; a 13B-parameter language model cannot, at least not with current proof systems (\S\ref{sec:zkml}). This tension between expressiveness and verifiability is the central challenge motivating our trust framework.

\subsection{Formal Model: Agent-Blockchain Interaction Model (ABIM)}\label{sec:abim}

To reason precisely about security properties across all layers of our framework, we define the Agent-Blockchain Interaction Model (\abim). The model serves two purposes: it provides a shared formal vocabulary for stating the guarantees that each layer must provide, and it makes explicit the assumptions under which those guarantees hold. We first define the model's components, then state the security properties that each layer must satisfy.

\subsubsection{Model Definition}

\abim is defined as a six-tuple:
\begin{equation}\label{eq:abim}
    \abim = \langle \mathcal{A}, \mathcal{B}, \mathcal{V}, \mathcal{P}, \mathcal{I}, \mathcal{E} \rangle
\end{equation}

\noindent The components are defined as follows.

\textbf{Agent Set $\mathcal{A}$.} $\mathcal{A} = \{a_1, a_2, \ldots, a_n\}$ is a finite set of agents. Each agent $a_i$ is a five-tuple:
\begin{equation}\label{eq:agent}
    a_i = (\mathit{id}_i,\; \mathit{keys}_i,\; \mathit{caps}_i,\; M_i,\; \alpha_i)
\end{equation}
where $\mathit{id}_i \in \mathbb{ID}$ is a unique on-chain identifier (e.g., an ERC-4337 account address), $\mathit{keys}_i \subseteq \mathbb{K}$ is a set of cryptographic keys the agent controls, $\mathit{caps}_i \subseteq \mathbb{C}$ is the set of declared capabilities (e.g., ``DeFi trading,'' ``governance voting''), $M_i$ is a reference to the agent's underlying AI model (which may be opaque), and $\alpha_i \in \{\textsc{assistive}, \textsc{semi-auto}, \textsc{auto}\}$ denotes the agent's autonomy level.

\textbf{Blockchain State Machine $\mathcal{B}$.} $\mathcal{B} = (S, s_0, T, \delta)$ is a deterministic state machine, where $S$ is the set of all valid blockchain states, $s_0 \in S$ is the genesis state, $T$ is the set of all well-formed transactions, and $\delta: S \times T \rightharpoonup S$ is a partial state transition function that maps a current state $s \in S$ and a valid transaction $t \in T$ to a successor state $s' = \delta(s, t)$, or is undefined ($\bot$) if $t$ is invalid in state $s$. Agents interact with $\mathcal{B}$ exclusively by submitting transactions; the model does not introduce a separate operation-level abstraction, since all agent actions must ultimately be encoded as transactions to produce on-chain state transitions.

\textbf{Verification Function $\mathcal{V}$.} $\mathcal{V}: T \times \Pi \to \{\texttt{accept}, \texttt{reject}\}$ determines whether a transaction is accompanied by a valid proof of correctness. Here $\Pi$ is the space of all proofs, which includes cryptographic proofs (e.g., ZK-SNARKs), fraud proofs (with associated challenge periods), hardware attestations (e.g., TEE remote attestation), and the null proof $\pi_\emptyset$ (no verification). The strength of the guarantee depends on the proof type, as formalized by the Trust Foundation properties below.

\textbf{Permission Policy $\mathcal{P}$.} $\mathcal{P}: \mathcal{A} \times T \times S \to \{\texttt{allow}, \texttt{deny}\}$ is a state-dependent permission function that determines whether agent $a$ is authorized to submit transaction $t$ in state $s$. The state dependence captures dynamic constraints such as session-key expiry, spending limits, and time-bounded delegation.

\textbf{Intent Function $\mathcal{I}$.} $\mathcal{I}: \mathit{NL} \to 2^{\mathit{Constraint}}$ maps a natural language specification $\mathit{nl} \in \mathit{NL}$ to a set of formal constraints $C = \{c_1, c_2, \ldots\}$, where each constraint $c_j$ is a predicate over post-execution states: $c_j: S \to \{\texttt{true}, \texttt{false}\}$. A solver $\mathit{solv}: 2^{\mathit{Constraint}} \times S \to T^*$ produces a transaction sequence $\bar{t} = (t_1, \ldots, t_k)$ that, when executed from state $s$, is intended to reach a state satisfying all constraints.

\textbf{Economic Incentive Function $\mathcal{E}$.} $\mathcal{E}: \mathcal{A} \times T \times S \times S \to \mathbb{R}$ assigns a real-valued reward (or penalty, if negative) to agent $a$ for submitting transaction $t$ that transitions the blockchain from state $s$ to state $s'$. This function captures staking rewards, slashing penalties, gas refunds, revenue-sharing distributions, and solver fees.

\subsubsection{Security Properties by Layer}\label{sec:abim-properties}

Using the components defined above, we state the security property that each framework layer must satisfy. These properties serve as formal specifications against which standards, protocols, and implementations can be evaluated; their satisfaction is not guaranteed by current systems, and the gap between the stated property and the state of the art motivates the open problems identified throughout this paper.

\textbf{Trust Foundation (TF): Verification Soundness.} The verification function must be sound: if it accepts a transaction with proof $\pi$, the transaction's claimed output must be correct.
\begin{equation}\label{eq:tf}
    \mathcal{V}(t, \pi) = \texttt{accept} \implies \mathit{output}(t) = \mathit{claimed}(t)
\end{equation}
The strength of this guarantee varies by proof type: cryptographic proofs (zkML) provide computational soundness under standard cryptographic assumptions; fraud proofs (opML) provide game-theoretic soundness assuming $\geq$1 honest challenger within the dispute window $\Delta$; hardware attestation (TEE) provides soundness conditional on the manufacturer's honesty.

\textbf{A1 (Identity): Uniqueness.} No two distinct agents may share the same on-chain identifier.
\begin{equation}\label{eq:a1}
    \forall a_i, a_j \in \mathcal{A}: a_i \neq a_j \implies \mathit{id}_i \neq \mathit{id}_j
\end{equation}
Non-forgeability, the requirement that no adversary can produce valid transactions under $\mathit{id}_i$ without controlling $\mathit{keys}_i$, is inherited from the underlying cryptographic signature scheme and is not restated here.

\textbf{A2 (Permission): Enforcement Completeness.} If the permission policy denies a transaction, the protocol must ensure that the transaction does not produce a state transition. This must hold even when the agent composes multiple individually permitted transactions.
\begin{equation}\label{eq:a2}
    \forall a \in \mathcal{A}, t \in T, s \in S: \mathcal{P}(a, t, s) = \texttt{deny} \implies \delta(s, t) = \bot
\end{equation}

\textbf{A3 (Intent): Solver Faithfulness.} For any intent with constraint set $C$ and solver-produced transaction sequence $\bar{t}$, executing $\bar{t}$ from the current state $s$ must reach a state $s'$ that satisfies every constraint.
\begin{equation}\label{eq:a3}
    \forall C = \mathcal{I}(\mathit{nl}),\; \bar{t} = \mathit{solv}(C, s): \bigwedge_{c \in C} c\bigl(\delta^*(s, \bar{t})\bigr) = \texttt{true}
\end{equation}
where $\delta^*(s, \bar{t}) = \delta(\cdots\delta(\delta(s, t_1), t_2)\cdots, t_k)$ denotes sequential execution of the transaction sequence.

\textbf{A4 (Economy): Incentive Compatibility.} The economic function must be incentive-compatible: an agent maximizing its expected reward should submit transactions that are beneficial to the system (or at minimum, non-harmful).
\begin{equation}\label{eq:a4}
    \forall a \in \mathcal{A}: \arg\max_{t} \mathbb{E}_{t_{-a}}[\mathcal{E}(a, t, s, s')] \in T_{\textit{beneficial}}
\end{equation}
where $T_{\textit{beneficial}} \subseteq T$ is the set of transactions that satisfy application-specific utility criteria (e.g., faithful model inference, honest governance voting), and the expectation $\mathbb{E}_{t_{-a}}$ is taken over the concurrent transactions of other agents, which determine the realized state $s' = \delta^*(s, \bar{t})$.

\textbf{B1 (Security): Detection Completeness.} An agent serving as a security auditor should detect all vulnerabilities in the target contract within the scope of its declared capabilities.
\begin{equation}\label{eq:b1}
    \forall v \in \mathit{Vuln}(s) \cap \mathit{scope}(a.\mathit{caps}): \Pr[\mathit{detect}(a, v) = \texttt{true}] \geq 1 - \epsilon
\end{equation}
where $\mathit{Vuln}(s)$ is the set of true vulnerabilities in state $s$, and $\epsilon$ is the model's inherent false-negative rate. In practice, current systems achieve $\epsilon$ values ranging from 0.02 (LLM-SmartAudit on benchmarks) to 0.40 (independent evaluations), as detailed in \S\ref{sec:b1}.

\textbf{B2 (Consensus): Validator Verifiability.} Every AI validator's consensus decision must be accompanied by a proof that the verification function accepts. Let $\mathit{Val} \subseteq \mathcal{A}$ denote the set of agents participating as validators. A validator's attestation is modeled as a transaction $t_a^{\textit{att}} \in T$ that encodes its decision $\mathit{dec}_a \in \{\texttt{commit}, \texttt{abort}\}$ and the proposed block, with accompanying proof $\pi_a \in \Pi$.
\begin{equation}\label{eq:b2}
    \forall a \in \mathit{Val}: \mathcal{V}(t_a^{\textit{att}}, \pi_a) = \texttt{accept}
\end{equation}
This property is in tension with the probabilistic nature of AI models: an honest AI validator with error rate $\epsilon_a$ may produce incorrect decisions with probability $\epsilon_a$, so the classical BFT assumption of deterministic correctness must be replaced by a probabilistic bound (\S\ref{sec:b2}).

\textbf{B3 (Governance): Bounded Influence.} The aggregate voting power of AI agents must not exceed a fraction $\tau$ of total voting power, ensuring that governance outcomes reflect the preferences of human stakeholders.
\begin{equation}\label{eq:b3}
    \sum_{a \in \mathcal{A}_{\textit{AI}}} w(a, s) \leq \tau \cdot \sum_{p \in \mathcal{A}_{\textit{all}}} w(p, s)
\end{equation}
where $w(a, s)$ is the voting weight of participant $a$ in state $s$ (which may be token-weighted, quadratic, or delegated), $\mathcal{A}_{\textit{AI}} \subseteq \mathcal{A}$ is the set of AI agent voters, and $\tau \in (0, 1)$ is a governance parameter. No current DAO enforces such a bound, as discussed in \S\ref{sec:b3}.

\subsubsection{Model Properties and Limitations}

The \abim formalization captures the essential structure of agent--blockchain interactions but deliberately abstracts away several aspects. First, it models blockchain state transitions as deterministic and sequential, omitting concurrency and mempool dynamics (which are relevant to MEV, addressed in \S\ref{sec:a3}). Second, the intent function $\mathcal{I}$ is assumed to produce well-formed constraint sets, whereas in practice, LLM-based intent parsing may produce ambiguous or contradictory constraints. Third, the model treats each agent as a single entity, whereas real agents may operate as multi-agent systems internally (as in LLM-SmartAudit's multi-agent auditing pipeline, \S\ref{sec:b1}). Fourth, the six-tuple is structurally biased toward the \dirA direction: $\mathcal{V}$, $\mathcal{P}$, $\mathcal{I}$, and $\mathcal{E}$ each map directly to a \dirA layer, whereas the \dirB properties (Eqs.~\ref{eq:b1}--\ref{eq:b3}) introduce auxiliary notation ($\mathit{Vuln}(s)$, $\mathit{detect}()$, $\mathit{Val}$, $w(a,s)$) that is not grounded in the tuple. This asymmetry reflects the current state of the field: the \dirA direction has mature infrastructure that admits compact formal characterization, while the \dirB direction remains exploratory and resists premature unification.

We also note that the eight security properties stated above are not of uniform type. Properties~\ref{eq:a1} and~\ref{eq:a2} are \emph{protocol invariants} that can, in principle, be enforced on-chain. Properties~\ref{eq:tf},~\ref{eq:b1}, and~\ref{eq:b2} are \emph{assurance bounds} whose strength depends on cryptographic assumptions or probabilistic error rates~$\epsilon$. Properties~\ref{eq:a3},~\ref{eq:a4}, and~\ref{eq:b3} are \emph{mechanism design objectives} that can only be evaluated through equilibrium analysis, not enforced by protocol logic alone. This three-way distinction matters when assessing gap severity in later sections: an unmet invariant is an engineering deficit, an unmet assurance bound calls for stronger proof systems, and an unmet mechanism objective requires new theoretical frameworks.

Despite these simplifications, the model is expressive enough to state the security properties that distinguish the layers of our framework and to identify where current systems fall short of the stated guarantees.

\subsection{Five-Dimensional Evaluation Framework}\label{sec:eval-framework}

To evaluate EIPs, ERCs, academic proposals, and industry projects on a common basis, we define a five-dimensional evaluation framework. Each dimension captures a distinct property that is essential for reasoning about the suitability of agent--blockchain infrastructure, and together they span the design trade-off space from cryptographic soundness to real-world deployability.

\begin{enumerate}[label=\textbf{D\arabic*.},leftmargin=2.5em]
\item \textbf{Verifiability} measures the strength of correctness guarantees. We distinguish four levels: \emph{High} (cryptographic proof, \eg zkML or ZKP-based verification, where correctness follows from mathematical assumptions), \emph{Medium} (economic or game-theoretic guarantee, \eg fraud proofs with staked collateral, where correctness holds under incentive-compatibility assumptions), \emph{Low} (reputation or attestation only, \eg a trusted registry or hardware attestation, where correctness depends on the honesty of a designated party), and \emph{None} (no on-chain verification mechanism).

\item \textbf{Minimality of Trust} counts the minimum number of trusted entities required beyond the blockchain protocol itself. A score of \emph{0} means fully trustless (only cryptographic assumptions); \emph{1} means one trusted party (\eg a TEE manufacturer or a single honest challenger in optimistic verification); \emph{2+} means multiple trusted parties (\eg a committee, a set of attesters, or an oracle network). Lower is better; this dimension captures how well the standard or system aligns with blockchain's trust-minimization ethos.

\item \textbf{Expressiveness} measures the range of agent behaviours or use cases that the standard can support. We rate this as \emph{Very High} (supports arbitrary agent operations including natural language interfaces and cross-chain coordination), \emph{High} (supports a broad class of agent operations within a single domain), \emph{Medium} (supports a restricted but useful set of operations), or \emph{Low} (supports only narrowly defined operations).

\item \textbf{Composability} measures how easily a standard can be combined with other standards or integrated into existing ecosystems. \emph{Very High} means the standard extends established primitives (\eg ERC-721, ERC-4337) and is designed for plug-and-play interoperability; \emph{High} means it defines clean interfaces that other standards reference; \emph{Medium} means it can interoperate with effort but is not designed for modular composition; \emph{Low} means it introduces proprietary or hardware-specific interfaces.

\item \textbf{Maturity} reflects the governance status of the standard as defined in the EIP process (\S\ref{sec:bg-accounts}): \emph{Final} (adopted, stable specification), \emph{Review} (under formal peer review), \emph{Draft} (initial proposal, interfaces may change), or \emph{Stagnant/Withdrawn} (inactive or retracted). For academic proposals and industry projects that do not follow the EIP process, we map their development stage to the closest equivalent.
\end{enumerate}

The first four dimensions capture inherent design properties that involve fundamental trade-offs: for example, stronger Verifiability typically comes at the cost of lower Expressiveness (proving correctness limits the complexity of supported operations), and higher Trust Minimality is often achieved by sacrificing Composability (trustless designs may require custom interfaces incompatible with existing infrastructure). Maturity is an orthogonal, temporal dimension that tracks progress toward stable specification rather than a design trade-off. In the evaluation tables throughout \S\ref{sec:trust-foundation}--\S\ref{sec:direction-b}, each standard is scored by the authors based on its specification, reference implementations, and (where available) third-party audits. Where scores involve judgment (\eg whether a standard's expressiveness is ``Medium'' or ``High''), we err on the side of the interpretation most consistent with the standard's stated scope.

\subsection{Threat Model}\label{sec:threat-model}

We identify seven categories of threats to agent--blockchain interactions (\Cref{tab:threats}). Three of them, agent impersonation, privilege escalation, and intent deception, target the \dirA layers and exploit the gap between what agents are authorized to do and what they actually do. Two others, inference tampering and model backdoors, target the Trust Foundation, undermining the correctness guarantees that all other layers depend on. Solver collusion and agent collusion target multi-party interactions where individually rational behaviour can produce collectively harmful outcomes. The ``Gap'' column in \Cref{tab:threats} highlights how many of these threats remain unaddressed by current standards and research.

\begin{table}[t]
\caption{Threat model for agent-blockchain interactions.}\label{tab:threats}
\centering
\small
\begin{tabular}{@{}llcl@{}}
\toprule
\textbf{Threat} & \textbf{Capability} & \textbf{Layer} & \textbf{Gap} \\
\midrule
Agent Impersonation & Forge identity & A1, B1 & No Sybil-resistant agent ID \\
Privilege Escalation & Exceed scope & A2 & No formal verification tools \\
Intent Deception & Unfaithful solver & A3 & Immature intent verification \\
Inference Tampering & Wrong AI result & TF & LLM verification infeasible \\
Solver Collusion & MEV extraction & A3 & Insufficient theory \\
Agent Collusion & Manipulate governance & B3 & Completely unexplored \\
Model Backdoor & LLM backdoor & TF, B2 & Completely unexplored \\
\bottomrule
\end{tabular}
\end{table}

%% ==========================================================
%% SECTION 3: TRUST FOUNDATION
%% ==========================================================
\section{Trust Foundation: Verifiable Computation}\label{sec:trust-foundation}

A central question that spans both the \dirA and \dirB directions is: \emph{how can one verify that an AI agent's reasoning and outputs are correct without re-executing the computation?} This is the verification soundness property stated in \Cref{eq:tf}: if the verification function $\mathcal{V}$ accepts an action with proof $\pi$, the action's output must be correct. Not every agent interaction requires cryptographic verification: a read-only analytics agent or a low-value assistive bot may operate under weaker trust models. However, as agent autonomy and transaction stakes increase, verifiable computation becomes increasingly critical.

Three families of techniques have emerged, which we organize by \emph{decreasing trust minimality} (or equivalently, increasing practical deployability). First, \emph{zkML} (\S\ref{sec:zkml}) achieves the strongest guarantee, cryptographic proofs that require zero trust assumptions beyond mathematics, but at the cost of extreme computational overhead that limits the scale of models it can verify. Second, \emph{opML} (\S\ref{sec:opml}) relaxes the requirement to a game-theoretic guarantee, assuming at least one honest challenger exists, in exchange for practical scalability to LLM-scale models. Third, \emph{TEE} (\S\ref{sec:tee}) shifts trust entirely to hardware manufacturers, achieving the best performance and broadest model support at the cost of a centralized trust assumption. This ordering, from strongest-but-slowest to weakest-but-fastest, defines the fundamental design trade-off that agent system architects must navigate. We then provide a \emph{comparative analysis} (\S\ref{sec:tf-comparison}) using our five-dimensional framework, followed by a discussion of the central open problem (\S\ref{sec:llm-bottleneck}).

\subsection{Zero-Knowledge Machine Learning (zkML)}\label{sec:zkml}

The line of work on zkML aims to generate succinct, non-interactive proofs that a particular model produced a particular output on a particular input, without revealing either the model weights or the input data.

The earliest practical results targeted convolutional networks. Liu~\etal~\cite{liu2021zkcnn} showed that the sumcheck protocol can be adapted to verify CNN predictions in time linear in the circuit size (zkCNN), proving VGG-16 (15M parameters) in 88.3 seconds with 341\,KB proofs. Around the same time, Weng~\etal~\cite{weng2021mystique} demonstrated that quantized networks as deep as ResNet-101 (over 100 layers) can be handled through efficient conversion protocols for subfield Vector Oblivious Linear Evaluation (sVOLE)-based ZK proofs (Mystique), achieving a 7$\times$ improvement in matrix multiplication proofs over prior work. These results established feasibility, but the models involved, image classifiers with millions of parameters, are orders of magnitude smaller than the language models that power contemporary agents.

Closing this gap, Sun~\etal~\cite{sun2024zkllm} proposed zkLLM, the first system capable of proving inference for a 13B-parameter language model. Two novel primitives, \texttt{tlookup} for non-arithmetic operations and \texttt{zkAttn} for the attention mechanism, bring proving time for the entire inference process (sequence length 2048) down to under 15 minutes. While this represents a milestone, such latency is still far too slow for interactive agent scenarios that demand sub-second responses.

Complementary work has focused on tooling and compilation. The ZKML compiler~\cite{kang2024zkml-system} achieves up to 5$\times$ faster proving and 22$\times$ smaller proofs compared to prior CNN-focused systems (zkCNN, vCNN), while extending support to production-scale models including GPT-2 and Twitter's recommendation pipeline. ZEN~\cite{feng2021zen} pioneered R1CS-friendly quantization, achieving 5.4--22.2$\times$ constraint reduction (15.4$\times$ average). zkPyTorch~\cite{zkpytorch2025} bridges PyTorch directly to ZKP engines, proving VGG-16 inference in 2.2 seconds per image and Llama-3 8B in 150 seconds per token. zkRNN~\cite{zkrnn2026} extends this line of work to recurrent architectures with polylogarithmic proof size and millisecond-scale verification. On the training side, Abbaszadeh~\etal~\cite{lee2024kaizen} introduced Kaizen for verifying DNN training via GKR-style interactive proofs with recursive proof composition (1.63\,MB proofs, constant verifier cost), while Garg~\etal~\cite{garg2023training-zk} provided early experimental validation of training proofs for logistic regression with streaming-friendly memory usage. Two on-chain standards complement this body of work: ERC-7992~\cite{erc7992} standardizes ML model registration and ZK inference proof verification, and ERC-7007~\cite{erc7007} defines verifiable AI-generated content NFTs. For broader context, several surveys map the ZKML landscape from complementary angles: Ker{\v{s}}i{\v{c}}~\etal~\cite{onchain-zkml2024} benchmark two leading on-chain frameworks (EZKL and Orion) with end-to-end deployment experiments, Peng~\etal~\cite{peng2025zkml-survey} catalogue 27 ZKML studies spanning training, inference, and testing, and Zhang~\etal~\cite{zkp-decentralized-ml2023} systematize 56 ZKP-VML schemes into three main technical routes (GKR/sumcheck, R1CS/circuit, and MPC-based).

\subsection{Optimistic Machine Learning (opML)}\label{sec:opml}

Rather than proving correctness upfront, opML borrows the optimistic paradigm from rollups: inference results are assumed correct unless challenged within a dispute window. Conway~\etal~\cite{conway2024opml} showed that this approach can support 7B-parameter LLaMA models running on standard PCs without GPU acceleration. The trade-off is latency: results are not finalized until the challenge period expires. So~\etal~\cite{so2024oppai} combined the privacy guarantees of zkML with the efficiency of opML in a hybrid framework (opp/ai), suggesting that the two approaches are more complementary than competing.

\subsection{Trusted Execution Environments (TEE)}\label{sec:tee}

TEEs, exemplified by Intel SGX and ARM TrustZone, provide hardware-attested execution in an isolated enclave. Tram\`{e}r and Boneh~\cite{tramer2019slalom} showed that inference can be split between a TEE enclave (for verification) and an untrusted GPU (for computation), achieving 6--20$\times$ throughput improvement for verifiable inference and 4--11$\times$ for private inference on VGG-16 and MobileNet. Subsequent work has tackled the primary bottleneck, SGX's limited Enclave Page Cache of roughly 90\,MB, through memory-efficient inference designs: \cite{tee-dl-memory2021} reduces VGG-16 inference overhead from 26$\times$ to only 1.09$\times$ via Y-plane partitioning, while InferONNX~\cite{papafragkaki2025inferonnx} achieves a compact 46\,MB runtime footprint with 1.5--4$\times$ faster inference via model partitioning. A survey by Chaudhuri~\etal~\cite{chaudhuri2025secured} provides a comprehensive overview of TEE implementations across CPU and GPU platforms, covering NVIDIA H100 confidential computing and related approaches. TEE is the approach most widely adopted in industry today (e.g., bAI Fund~\cite{baifund2025} executes agent fund operations entirely within TEE enclaves on Morph L2), but its trust assumption, that the hardware manufacturer's attestation is honest, sits uncomfortably with blockchain's trust-minimization ethos. An alternative trust paradigm, fully homomorphic encryption (FHE), enables computation on encrypted data without decryption; Mind Network~\cite{mindnetwork2024} is among the few projects in our survey that adopt FHE as a core trust primitive through its HTTPZ protocol, providing quantum-resistant guarantees but with significant computational overhead.

\subsection{Comparative Analysis}\label{sec:tf-comparison}

\begin{figure}[t]
\centering
\begin{tikzpicture}
\footnotesize
%% --- Helper: draw one radar background ---
%% We use manual polar coordinates for a 5-axis radar
%% Axes: Verifiability(90), TrustMin(162), Expressiveness(234), Composability(306), Maturity(18)
%% Angles for 5 axes: 90, 162, 234, 306, 378(=18)

%% Axis angles
\def\anga{90}   % Verifiability
\def\angb{162}  % Trust Minimality
\def\angc{234}  % Expressiveness
\def\angd{306}  % Composability
\def\ange{18}   % Maturity

%% Scale: 1 unit = 0.6cm, max = 5
\def\rscale{0.55}

%% --- Draw grid ---
\foreach \r in {1,2,3,4,5} {
    \draw[black!15, thin]
        ({\anga}:{\r*\rscale}) --
        ({\angb}:{\r*\rscale}) --
        ({\angc}:{\r*\rscale}) --
        ({\angd}:{\r*\rscale}) --
        ({\ange}:{\r*\rscale}) -- cycle;
}
%% --- Draw axes ---
\foreach \ang in {\anga,\angb,\angc,\angd,\ange} {
    \draw[black!30, thin] (0,0) -- ({\ang}:{5*\rscale});
}
%% --- Axis labels ---
\node[above, font=\scriptsize\bfseries] at ({\anga}:{5*\rscale+0.35}) {Verifiability};
\node[left, font=\scriptsize\bfseries] at ({\angb}:{5*\rscale+0.35}) {Trust Minimality};
\node[below left, font=\scriptsize\bfseries] at ({\angc}:{5*\rscale+0.35}) {Expressiveness};
\node[below right, font=\scriptsize\bfseries] at ({\angd}:{5*\rscale+0.35}) {Composability};
\node[right, font=\scriptsize\bfseries] at ({\ange}:{5*\rscale+0.35}) {Maturity};
%% --- Scale labels ---
\foreach \r/\lbl in {1/1,3/3,5/5} {
    \node[font=\tiny, black!40, anchor=south west] at ({\anga}:{\r*\rscale}) {\lbl};
}

%% --- zkML: Verif=5, TrustMin=5, Express=2, Compos=4, Maturity=2 ---
\draw[blue!70, thick, fill=blue!15, fill opacity=0.4]
    ({\anga}:{5*\rscale}) --
    ({\angb}:{5*\rscale}) --
    ({\angc}:{2*\rscale}) --
    ({\angd}:{4*\rscale}) --
    ({\ange}:{2*\rscale}) -- cycle;
\foreach \ang/\val in {\anga/5, \angb/5, \angc/2, \angd/4, \ange/2} {
    \fill[blue!70] ({\ang}:{\val*\rscale}) circle (1.5pt);
}

%% --- opML: Verif=3, TrustMin=3, Express=4, Compos=3, Maturity=3 ---
\draw[green!60!black, thick, fill=green!15, fill opacity=0.35]
    ({\anga}:{3*\rscale}) --
    ({\angb}:{3*\rscale}) --
    ({\angc}:{4*\rscale}) --
    ({\angd}:{3*\rscale}) --
    ({\ange}:{3*\rscale}) -- cycle;
\foreach \ang/\val in {\anga/3, \angb/3, \angc/4, \angd/3, \ange/3} {
    \fill[green!60!black] ({\ang}:{\val*\rscale}) circle (1.5pt);
}

%% --- TEE: Verif=2, TrustMin=2, Express=5, Compos=2, Maturity=5 ---
\draw[red!70, thick, fill=red!12, fill opacity=0.35]
    ({\anga}:{2*\rscale}) --
    ({\angb}:{2*\rscale}) --
    ({\angc}:{5*\rscale}) --
    ({\angd}:{2*\rscale}) --
    ({\ange}:{5*\rscale}) -- cycle;
\foreach \ang/\val in {\anga/2, \angb/2, \angc/5, \angd/2, \ange/5} {
    \fill[red!70] ({\ang}:{\val*\rscale}) circle (1.5pt);
}

%% --- Legend ---
\node[font=\scriptsize] at (0, -3.5) {
    \tikz{\draw[blue!70, thick, fill=blue!15, fill opacity=0.4] (0,0) rectangle (0.35,0.2);} zkML \quad
    \tikz{\draw[green!60!black, thick, fill=green!15, fill opacity=0.35] (0,0) rectangle (0.35,0.2);} opML \quad
    \tikz{\draw[red!70, thick, fill=red!12, fill opacity=0.35] (0,0) rectangle (0.35,0.2);} TEE
};

\end{tikzpicture}
\caption{Five-dimensional radar chart comparing verification approaches. zkML achieves the highest verifiability and trust minimality but lags in expressiveness and maturity. TEE excels in expressiveness and maturity but requires hardware trust. opML provides a balanced middle ground. Scores range from 1 (worst) to 5 (best).}
\label{fig:radar}
\end{figure}
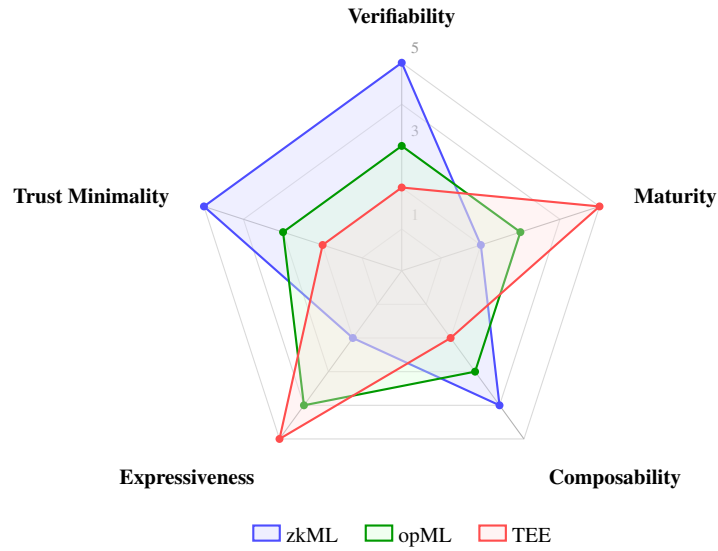

\Cref{fig:radar} visualizes the three approaches along our five dimensions (defined in \S\ref{sec:eval-framework}). The scores are assigned as follows. \emph{Verifiability}: zkML scores 5 (cryptographic, non-interactive proof), opML scores 3 (probabilistic, contingent on a challenger appearing), TEE scores 2 (hardware attestation, no cryptographic guarantee of computation correctness). \emph{Trust Minimality}: zkML requires 0 trusted parties (only mathematical assumptions), opML requires 1 (at least one honest challenger), TEE requires 1 (the hardware manufacturer). \emph{Expressiveness}: TEE scores 5 (supports any model on commodity hardware), opML scores 4 (demonstrated on 7B-parameter LLMs~\cite{conway2024opml}), zkML scores 2 (limited to small-to-medium models, with LLM-scale proofs requiring $\sim$15 minutes for a full inference pass (\S\ref{sec:zkml})). \emph{Composability}: zkML scores 4 (standard proof formats integrate with on-chain verifiers via ERC-7992), opML scores 3 (challenge-period delays complicate synchronous composition), TEE scores 2 (hardware-specific attestation formats limit portability). \emph{Maturity}: TEE scores 5 (deployed in production, e.g., bAI Fund~\cite{baifund2025}), opML scores 3 (early deployments, e.g., OPML on Optimism), zkML scores 2 (research-stage with limited production use).

The comparison points to a single dominant tension rather than a clean winner. zkML and TEE anchor opposite ends of the trust-performance spectrum, and opML sits between them on every dimension, so the choice of verification approach depends on the application context rather than on any intrinsic superiority. This non-domination is precisely what makes hybrid architectures attractive as the near-term path: opp/ai~\cite{so2024oppai} shows that partitioning a model between zkML (for privacy-sensitive submodels) and opML (for efficiency-sensitive submodels) can balance privacy and computational cost, and the same principle could extend to TEE--zkML combinations in which TEE handles latency-sensitive operations while zkML provides asynchronous auditability. What makes the situation precarious is the maturity gap between TEE (deployed) and zkML (research): industry is converging on the weakest trust model by default, and if TEE attestation is compromised, as has happened with Intel SGX side-channel attacks, the entire agent infrastructure built on it loses its security foundation. From an \abim perspective, verification soundness (\Cref{eq:tf}) is an assurance bound whose strength depends on the proof type: zkML provides computational soundness but only for models up to $\sim$1B parameters; opML provides game-theoretic soundness contingent on challenger availability; TEE provides hardware-conditional soundness. At LLM scale, no current approach satisfies \Cref{eq:tf} with guarantees strong enough for high-stakes autonomous operations.

%% ===== Fig 3: Five-Dimensional Radar Charts =====

% \begin{table}[t]
% \caption{Five-dimensional comparison of verification approaches.}
% \label{tab:tf-comparison}
% \centering
% \small
% \begin{tabular}{lccc}
% \toprule
% \textbf{Dimension} & \textbf{zkML} & \textbf{opML} & \textbf{TEE} \\
% \midrule
% Verifiability & Cryptographic & Probabilistic & Hardware attestation \\
% Trust Minimality$^\dagger$ & 0 & 1 & 1 \\
% Expressiveness & Limited & High & Full \\
% Composability & High & Medium & Low \\
% Maturity & Research & Early Deploy & Deployed \\
% \bottomrule
% \end{tabular}
% \\[2pt]
% \footnotesize $^\dagger$Trust Minimality indicates the number of trusted parties required: 0 = trustless; 1 = one trusted assumption (e.g., an honest challenger or the hardware vendor).
% \end{table}

\subsection{OP1: LLM Verifiability Bottleneck}\label{sec:llm-bottleneck}

The central open problem is what we term the \emph{LLM verifiability bottleneck}. As shown in \S\ref{sec:zkml}, the most capable zkML system to date requires roughly 15 minutes to prove a full inference pass for a 13B-parameter model, while agent-facing applications demand sub-second responses. Recent compiler advances narrow this gap: zkPyTorch~\cite{zkpytorch2025} reduces Llama-3 8B proving to 150 seconds per token, and zkRNN~\cite{zkrnn2026} achieves second-scale proving for recurrent architectures, but orders-of-magnitude improvements are still needed. Peng~\etal~\cite{peng2025zkml-survey} survey 27 ZKML studies and organize them into verifiable training, testing, and inference, while the end-to-end verifiable AI framework of~\cite{e2e-verifiable-ai2025} identifies a critical gap: no existing work connects proofs of training to proofs of inference into a single trust chain. Related systematization efforts include a SoK on verifiable federated learning~\cite{sok-verifiable-fl2025} and a survey on privacy-preserving LLM inference~\cite{llm-privacy-survey2026}, which identifies TEE-based solutions as deployable today while crypto-augmented designs (TEE + MPC/FHE) can reduce hardware trust in the medium term. Both efforts highlight the gap between theoretical feasibility and practical deployment. Promising directions include hierarchical proofs that verify only critical reasoning steps, approximate proofs that trade a small soundness gap for dramatic speedups, and hybrid TEE--ZKP architectures that use hardware attestation for latency-sensitive operations while generating cryptographic proofs asynchronously. Until this bottleneck is substantially reduced, the practical deployment of verifiable autonomous agents will remain constrained to the opML and TEE trust models.

%% ==========================================================
%% SECTION 4: DIRECTION A
%% ==========================================================
\section{B\,$\boldsymbol{\rightarrow}$\,A: Blockchain as Trust Infrastructure for Agents}\label{sec:direction-a}

\subsection{A1: Account and Identity}\label{sec:a1}

The first question any agent faces on-chain is: \emph{how does it establish a verifiable presence?} Answering this requires addressing three progressively deeper sub-problems. First, the agent needs an \emph{account mechanism} that supports programmable validation and delegated signing, since legacy EOAs lack these capabilities. Second, given such an account, the agent needs to be \emph{discoverable and assessable} by users and other agents, which requires identity and reputation standards. Third, the system must ensure that each agent identity is \emph{unique and non-duplicable}, preventing Sybil attacks at the agent layer. These three sub-problems form a logical progression: account infrastructure $\to$ identity semantics $\to$ uniqueness guarantees.

\subsubsection{Account Abstraction: The Foundation}

As described in \S\ref{sec:bg-accounts}, the evolution from EOA to ERC-4337~\cite{erc4337} to EIP-7702~\cite{eip7702} provides progressively more flexible account infrastructure for AI agents. Meta-transaction precursors such as ERC-2771~\cite{erc2771} first established the relayer/forwarder pattern that allows accounts without ETH to interact via gas-paying intermediaries, but lacked native protocol support and required each recipient contract to opt in by trusting specific forwarders; ERC-4337 generalizes this idea into a protocol-level account abstraction with bundlers and paymasters. This layer must enforce the identity uniqueness property stated in \Cref{eq:a1}. Here we focus on the security implications and supporting standards not covered in the background.

EIP-3074~\cite{eip3074}, the predecessor to EIP-7702, introduced AUTH and AUTHCALL opcodes but was withdrawn due to security concerns around unrestricted delegation. Supporting infrastructure standards round out the A1 stack: ERC-7562~\cite{erc7562} restricts the EVM operations permitted during validation to prevent denial-of-service attacks on the mempool, EIP-5792~\cite{eip5792} and ERC-7821~\cite{erc7821} standardize batch call APIs and minimal executor interfaces, ERC-7766~\cite{erc7766} enables signature aggregation across multiple UserOperations to reduce gas costs, ERC-7836~\cite{erc7836} provides session-key-based call preparation, and ERC-7677~\cite{erc7677} defines a standardized Paymaster web service. Prior SoKs on cryptocurrency wallets~\cite{sok-crypto-wallets2024} and wallet security evaluations~\cite{wallet-security2020}, which found that on average 10.2\% of on-chain wallet contracts are vulnerable, provide useful baselines for assessing the security of agent-facing account designs.

\subsubsection{Agent Identity Standards}

The three agent identity standards introduced in \S\ref{sec:bg-accounts}, namely ERC-8004~\cite{erc8004} (discovery), ERC-8126~\cite{erc8126} (self-registration and verification with ZKP-backed privacy), and ERC-7857~\cite{erc7857} (agent-as-NFT with encrypted metadata), address complementary aspects of the identity problem. Beyond on-chain standards, academic work explores off-chain identity delegation: Saavedra~\cite{agent-identity-delegation2026} proposes a ``Delegation Tetrahedron'' extending self-sovereign identity with a Delegate role for cryptographically anchored authority transfer, while Rodriguez Garzon~\etal~\cite{agent-did-vc2025} demonstrate DID/VC-based mutual authentication for LLM agents, though their evaluation reveals that agents occasionally agreed to skip authentication entirely, highlighting the fragility of delegating security-critical orchestration to LLMs.

\subsubsection{Sybil Resistance for Agents}

Proof-of-Personhood protocols~\cite{pop-redemocratizing2017,pop-sybil-review2020} were designed to ensure one-person-one-account guarantees, but extending them to AI agents raises a different set of questions: an agent's ``uniqueness'' must be defined in terms of its model, operator, and declared capabilities rather than biological identity. ERC-4361~\cite{erc4361} provides Ethereum-based sign-in for off-chain services, and ERC-5573~\cite{erc5573} extends this with capability-based ReCaps, but neither was designed with non-human agents in mind.

\subsubsection{Synthesis and Evaluation}

\Cref{tab:a1-eval} evaluates A1 standards along the five dimensions defined in \S\ref{sec:eval-framework}. Three patterns emerge. First, there is a clear \emph{maturity inversion}: the foundational account abstraction standards (ERC-4337 at Review, EIP-7702 at Final) are far more mature than the agent-specific identity standards (ERC-8004 at Draft, ERC-8126 at Review), meaning that agents can already obtain flexible accounts but cannot yet register or discover each other through stable, finalized interfaces. Second, a \emph{verifiability--expressiveness trade-off} is visible: ERC-8126 achieves High verifiability through ZKP-based proofs but supports only a restricted set of identity attributes (Medium expressiveness), whereas ERC-8004 offers richer discovery semantics (Medium expressiveness with registry-based search) at the cost of Low verifiability, relying on off-chain attestations with no on-chain proof of correctness. Third, \emph{composability is generally high} across all five standards because each builds on established Ethereum primitives (ERC-721, ERC-4337, EIP-712 signatures), but the two agent identity standards (ERC-8004, ERC-8126) are not yet interoperable with each other, lacking a shared identity schema that would allow agents registered under one standard to be discovered through the other.

These observations point to a concrete gap: the A1 layer currently provides the \emph{plumbing} (account abstraction) but not the \emph{semantics} (stable identity standards) needed for agents to operate as first-class on-chain entities. Until at least one agent identity standard reaches Final status with broad adoption, agent discovery will remain ad hoc and platform-specific, limiting the composability that the \dirA direction requires. In \abim terms, identity uniqueness (\Cref{eq:a1}) is a protocol invariant that should be enforced on-chain. ERC-8126's ZKP-based registration partially instantiates this property for agents that voluntarily register, but without a Sybil-resistant mechanism, an adversary can create multiple identities, circumventing the invariant. This gap propagates downstream: bounded influence (\Cref{eq:b3}) in governance cannot be enforced if identity uniqueness is not assured.

\begin{table}[t]
\caption{Five-dimensional evaluation of A1 (Account \& Identity) standards.}\label{tab:a1-eval}
\centering
\small
\begin{tabular}{@{}lccccc@{}}
\toprule
\textbf{Standard} & \textbf{Verifiability} & \textbf{Trust} & \textbf{Expressiveness} & \textbf{Composability} & \textbf{Maturity} \\
\midrule
ERC-4337 & Med & 2 & High & High & Review \\
EIP-7702 & Med & 1 & High & High & Final \\
ERC-8004 & Low & 3+ & Med & High & Draft \\
ERC-8126 & High & 1 & Med & Med & Review \\
ERC-7857 & High & 1 & High & Med & Final \\
\bottomrule
\end{tabular}
\end{table}

\subsubsection{OP9: Sybil-Resistant Agent Identity}
A stable, uniquely-identifying registration mechanism is only half of what \Cref{eq:a1} demands. The harder half is Sybil resistance: preventing a single operator from spawning many agents that appear, to the protocol, as independent entities. Proof-of-Personhood protocols~\cite{pop-redemocratizing2017,pop-sybil-review2020} address this for humans by anchoring uniqueness to biometric or social-graph proofs, but neither approach transfers to AI agents, whose ``personhood'' is not tied to any biological substrate and whose capabilities can be duplicated at near-zero marginal cost. The problem is to define an agent-appropriate notion of uniqueness (perhaps grounded in model fingerprints, operator attestations, or staked collateral) and to design a registration protocol that enforces it without centralizing trust in a single registrar. This problem has broad downstream impact: both governance bounds (\Cref{eq:b3}, targeted by OP3) and legal accountability (OP8) presuppose that AI voters can be counted as distinct entities, a guarantee that no current standard provides.

\subsection{A2: Permission and Delegation}\label{sec:a2}

Once an agent has an on-chain identity (A1), the next question is: \emph{what is the agent permitted to do, and how are those permissions enforced?} We decompose this into three sub-problems that move from theory to practice. First, we review the \emph{theoretical foundations} of capability-based access control, which provides the formal underpinning for reasoning about least-privilege and delegation. Second, we examine how these ideas are instantiated in \emph{modular smart contract account} standards, which decompose account logic into composable, auditable modules. Third, we identify the gap between general-purpose delegation standards and the specific requirements of \emph{AI agent delegation}, where the delegate is a probabilistic system rather than a deterministic script. This progression from theory $\to$ general standards $\to$ agent-specific gap reveals where existing infrastructure falls short.

\subsubsection{Theoretical Foundations: Capability-Based Security}

When an agent operates on behalf of a user, the central security question is how to bound what the agent is permitted to do. This layer must satisfy the enforcement completeness property of \Cref{eq:a2}: denied actions must be non-executable, even under composition. Capability-based access control (CapBAC), originating from Dennis and Van Horn~\cite{dennis1966capbac}, provides the natural theoretical foundation: rather than maintaining access control lists, capabilities are unforgeable tokens that grant the bearer specific permissions. Nakamura~\etal~\cite{nakamura2020capbac} adapt this model to smart contracts, achieving more fine-grained delegation than prior work by allowing subjects to independently delegate individual action tokens to and from multiple parties. Xu~\etal~\cite{xu2018blendcac} introduce federated authorization delegation with BlendCAC for IoT scenarios, demonstrating a decentralized, scalable solution feasible for resource-constrained devices. Ali~\etal~\cite{ali2019baci} extend this to support both event-based and query-based delegation with SPIN model checking under LTL properties. Zhang~\etal~\cite{zhang2018sc-access-control} propose a three-contract framework (access control, judge, and register contracts) with both static policy validation and dynamic behaviour monitoring. Gao~\etal~\cite{multihop-delegation2022} address multi-hop delegation in the context of electronic health record sharing, using ciphertext-policy attribute-based proxy re-encryption (CP-ABRPE) with smart contracts that enforce a controllable maximum delegation depth at each hop. Sherazi~\etal~\cite{sherazi2022emergency-delegation} tackle a complementary problem, emergency delegation, where standard authorization policies must be temporarily overridden based on contextual constraints. Zero-trust frameworks~\cite{zerotrust-blockchain2025} further inform agent permission design, reporting 14\% encryption efficiency improvements and 18\% reduction in data access latency compared to traditional methods.

\subsubsection{Modular Smart Contract Accounts}

The current standards landscape reflects an ongoing competition between modular account architectures. ERC-7579~\cite{erc7579} decomposes account functionality into Validator, Executor, and Fallback Handler modules that can be independently installed, upgraded, and audited. ERC-6900~\cite{erc6900} offers an alternative modular account architecture that decomposes functionality into Validation, Execution, and Hook modules, enabling features such as session keys and spending limits to be installed as composable modules rather than built into the account itself. ERC-7780~\cite{erc7780} pushes modularity further by separating Policy, Signer, and Stateless Validator modules, giving developers fine-grained control over which components handle which aspects of account security. Underpinning these designs, ERC-7484~\cite{erc7484} provides a module registry that allows third parties to attest to the security of individual modules.

\subsubsection{Agent-Specific Delegation}

Despite the proliferation of modular account standards, few address agent delegation directly. ERC-7710~\cite{erc7710} is among the few that explicitly lists AI agents as a motivating use case. ERC-5639~\cite{erc5639} provides delegation at the wallet, contract, and token level, ERC-7741~\cite{erc7741} enables operator authorization through EIP-712 signatures supporting meta-transactions and cross-chain scenarios, while ERC-7715~\cite{erc7715} defines APIs for requesting permissions, but none addresses the specific trust model required when the delegate is a probabilistic AI system rather than a deterministic script.

A Quantstamp audit~\cite{quantstamp2024modular-audit} illustrates the practical risks: auditors discovered two high-severity vulnerabilities: session keys with ERC-20 spend limits could drain the MSCA by exploiting pre-execution hook assumptions about unchanged state across multiple calls, and incorrect storage key derivation caused all session keys to share the same permissions rather than having individual scopes. This finding underscores that modular flexibility comes at the cost of implementation complexity, and that no formal verification tools currently exist for reasoning about the composition of agent permission policies.

\Cref{tab:a2-eval} evaluates A2 standards along the five dimensions. The most striking finding is the \emph{universal immaturity}: every standard except ERC-5639 (Review) remains at Draft, and none has reached Final, making this the least mature layer in the entire framework. Verifiability is uniformly Medium or Low, reflecting the absence of formal verification tools for reasoning about module composition, a gap directly illustrated by the Quantstamp audit findings. Expressiveness varies substantially: ERC-7780 achieves Very High expressiveness through its fine-grained separation of Policy, Signer, and Stateless Validator modules, while ERC-7710, which lists AI agents as a motivating use case, offers only Medium expressiveness. This contrast highlights a recurring tension: the standards designed for maximum flexibility were not designed with agents in mind, while the one standard that addresses agents does so narrowly. In \abim terms, enforcement completeness (\Cref{eq:a2}) requires that a denied transaction never produces a state transition, even under composition. The Quantstamp audit finding, a spending-limit bypass through specific call sequences, demonstrates a concrete violation of this invariant. Until formal verification tools can prove properties about composed module interactions, the A2 layer will remain the weakest link in the \dirA trust chain.

\subsubsection{OP2: Formal Verification of Agent Permission Policies}
Given a permission policy $\mathcal{P}$ and a set of declared agent capabilities, can one formally verify that the policy prevents privilege escalation and enforces least privilege? This question extends classical capability-based security theory to the setting of composable smart contract modules, where multiple independently developed modules interact through shared state. The Quantstamp audit finding, a spending limit bypass through specific call sequences, illustrates that informal reasoning about module composition is insufficient. Promising directions include adapting SPIN model checking~\cite{ali2019baci} to ERC-7579 module graphs, and developing type-level guarantees that prevent cross-module state interference.

\begin{table}[t]
\caption{Five-dimensional evaluation of A2 (Permission \& Delegation) standards.}\label{tab:a2-eval}
\centering
\small
\begin{tabular}{@{}lccccc@{}}
\toprule
\textbf{Standard} & \textbf{Verifiability} & \textbf{Trust} & \textbf{Expressiveness} & \textbf{Composability} & \textbf{Maturity}\\
\midrule
ERC-7579 & Med & 1 & High & High & Draft \\
ERC-6900 & Med & 1 & High & Med & Draft \\
ERC-7710 & Low & 1 & Med & High & Draft \\
ERC-7780 & Med & 1 & Very High & High & Draft \\
ERC-5639 & Low & 1 & Med & High & Review \\
ERC-7715 & Low & 1 & Med & Med & Draft \\
\bottomrule
\end{tabular}
\end{table}

\subsection{A3: Intent Specification and Execution}\label{sec:a3}

With identity (A1) and permissions (A2) in place, the agent must \emph{express what it wants to achieve} and have that goal faithfully executed. We examine four aspects of this layer, following the lifecycle of an intent from concept to analysis. First, we introduce the \emph{intent paradigm} itself, explaining how it differs from traditional transaction construction and why it is particularly well-suited to AI agents. Second, we survey the \emph{ERC standards} that define the on-chain intent infrastructure. Third, we review the \emph{formal analysis} of intent market structure and MEV dynamics, which reveals fundamental incentive challenges. Fourth, we identify the distinct \emph{roles AI agents play} in the intent ecosystem as both generators and solvers. This structure moves from concept $\to$ implementation $\to$ theory $\to$ application, reflecting the current state of a paradigm that is deployed before it is fully understood.

\subsubsection{The Intent Paradigm}

The intent paradigm inverts the traditional model of transaction construction. Instead of specifying exact calldata, namely which operation to call, on which contract, and with which parameters, users declare \emph{desired outcomes}, and a competitive market of \emph{solvers} races to find execution paths that satisfy those constraints. The solver faithfulness property (\Cref{eq:a3}) requires that any solver-produced execution must reach a state satisfying all original intent constraints. This inversion is particularly significant for AI agents: an LLM can naturally express a high-level goal (``swap token~A for at least~$X$ of token~B before deadline~$D$'') but would struggle to construct the optimal sequence of contract calls across liquidity pools, bridges, and aggregators. Nagesh~\cite{nagesh2023demystifying} provides an industry analysis of the intent-centric thesis, identifying fragmented intent pools as a key barrier and AI as a critical enabler for handling intent complexity.

\subsubsection{Intent Standards}

Four ERC standards currently define the on-chain intent infrastructure. ERC-7521~\cite{erc7521} introduces the \texttt{UserIntent} object and allows any solver to participate in combining and executing intents. ERC-7683~\cite{erc7683} extends this design to cross-chain operations, defining a ``filler'' network for settlement across domains. ERC-7806~\cite{erc7806} unifies the roles of relayer, paymaster, and bundler into a single solver entity (building on EIP-7702) with an open execution model where any solver can participate. ERC-7845~\cite{erc7845} standardizes a universal orchestrator RPC endpoint that, among other use cases, can serve AI-driven systems by providing a structured JSON-RPC interface for intent submission.

\subsubsection{Formal Analysis of Intent Markets}

Academic analysis of intent markets remains sparse. The only formal treatment we are aware of, by Chitra~\etal~\cite{chitra2024intent-markets}, models execution costs as barriers to entry and proves that the equilibrium number of solver entrants $k^*$ is asymptotically $O(\sqrt{n})$ for exponential price distributions and $O(n^{1/3})$ for uniform distributions, producing oligopolistic structures. In a related setting, Ma~\etal~\cite{ma2025ofa-pbs} establish the existence of Nash equilibria in order flow auctions under proposer-builder separation and show that the more capable builder earns disproportionately higher revenue, a driver of builder centralization. A sharper negative result comes from Bahrani, Garimidi, and Roughgarden~\cite{bahrani2024mev-tfm}, who prove that no non-trivial transaction fee mechanism can simultaneously satisfy DSIC (for users) and BPIC (for block producers) when producers are active MEV extractors. Against this backdrop, Rasheed~\etal~\cite{shapley-mev2025} use Shapley values to achieve fair MEV redistribution, with polynomial-time computability when searcher valuations are additive.

Complementary work examines execution quality and mitigation. Bachu~\etal~\cite{bachu2025price-improvement} quantify that Dutch auction-based OFAs (UniswapX, 1inch Fusion) provide statistically significant price improvements averaging 4--5 basis points over on-chain-only routing, while Watts~\etal~\cite{failure-costs2025} propose ``failure costs'' as a mechanism whose asymmetric payoff structures ultimately benefit users. RediSwap~\cite{rediswap2025} contributes a provably DSIC, individually rational, and Sybil-proof MEV redistribution mechanism for constant-function market makers, and a parallel game-theoretic analysis~\cite{mev-game-theory2025} shows that searcher competition follows Bertrand-style dynamics, creating a prisoner's dilemma with empirically observed bid-to-value ratios of 0.889. The broader DeFi context is captured by Werner~\etal~\cite{werner2021sok-defi}, whose foundational SoK situates MEV within the decentralized finance landscape. Despite this progress, no work has analyzed the specific dynamics that arise when \emph{AI agents}, rather than human traders, generate and solve intents.

\subsubsection{AI Agent Roles}

We identify two distinct roles that agents can play in the intent ecosystem. As \emph{intent generators}, agents translate natural language goals into structured intent specifications via $\mathcal{I}$, relieving users of the need to understand protocol-specific semantics. As \emph{solvers}, agents compete to find optimal execution paths, leveraging their ability to process large volumes of on-chain state and market data. In practice, Olas~\cite{olas2024}, the most technically mature agent infrastructure (\$13.8M funding, 4000+ deployed agents across 8+ chains), coordinates agent instances via Tendermint BFT consensus running finite state machine applications, with Gnosis Safe multisig for on-chain execution. Alternative architectures include AgentFi's~\cite{agentfi2024} ``agent-as-NFT'' design that embeds smart contract wallets within NFTs (inspired by ERC-6551), and Maiga's~\cite{maiga2025} hybrid approach combining the ElizaOS framework with centralized GPT-4o inference for DeFi trading on BNB Chain. Neither role has been systematically studied in the academic literature.

\Cref{tab:a3-eval} reveals that all four intent standards remain at Draft, making A3 the only layer where no standard has progressed beyond the initial proposal stage. Expressiveness is the strongest dimension (High to Very High), reflecting the deliberate design goal of supporting flexible outcome specifications; ERC-7683 and ERC-7845 achieve Very High expressiveness by supporting cross-chain settlement and natural language input, respectively. However, Verifiability is uniformly Medium or Low: no standard provides cryptographic proof that a solver's execution faithfully satisfies the original intent, instead relying on economic incentives (staked collateral, competitive auctions) that create game-theoretic rather than mathematical guarantees. This gap is compounded by the formal analysis: oligopolistic solver markets~\cite{chitra2024intent-markets} and the fundamental impossibility of simultaneously incentive-compatible fee mechanisms~\cite{bahrani2024mev-tfm} suggest that the economic security assumptions underlying current intent architectures may be weaker than assumed. Solver faithfulness (\Cref{eq:a3}) is a mechanism design objective in our \abim classification: it cannot be enforced by protocol logic alone but must be approximated through incentive alignment. The absence of any standard providing cryptographic proof that a solver's execution satisfies the original constraints means that \Cref{eq:a3} currently rests entirely on economic penalties rather than mathematical guarantees.

\begin{table}[t]
\caption{Functional and five-dimensional evaluation of A3 (Intent \& Execution) standards.}\label{tab:a3-eval}
\centering
\small
\setlength{\tabcolsep}{4pt}
\begin{tabular}{@{}lllcccccc@{}}
\toprule
\textbf{Standard} & \textbf{Intent Type} & \textbf{Agent Role} & \textbf{Verifiability} & \textbf{Trust} & \textbf{Expressiveness} & \textbf{Composability} & \textbf{Maturity} \\
\midrule
ERC-7521 & General & Generator/Solver & Med & 2 & High & High & Draft \\
ERC-7683 & Cross-chain & Solver & Med & 2 & Very\ High & High & Draft \\
ERC-7806 & EOA-based & Solver & Med & 1 & High & High & Draft \\
ERC-7845 & NL / voice & Generator & Low & 2 & Very\ High & Med & Draft \\
\bottomrule
\end{tabular}
\end{table}

\subsubsection{OP6: End-to-End Intent-to-Proof Pipeline}
The sharpest research opportunity at this layer is making the full pipeline from natural language intent to on-chain execution verifiable. This would require composing several individually challenging components: natural language understanding (translating user goals to formal specifications, perhaps through languages like DAOLang~\cite{agentdao2025}), solver execution (finding and executing optimal paths), and proof generation (certifying that the execution satisfied the original intent). No existing system addresses more than one of these steps.

\subsubsection{OP5: Inter-Agent Game Theory}
A second open problem concerns the multi-agent dynamics that intent markets create. When multiple AI agents compete on-chain, bidding in MEV auctions, solving intents, and voting in governance, standard game-theoretic analysis assumes rational players with known utility functions. AI agents introduce additional complexity: their strategies are learned rather than designed, their utility functions may be misaligned with their operators' intentions, and they can adapt their strategies in real time. Multi-agent reinforcement learning in blockchain environments and mechanism design for AI solver markets are two promising directions.

\subsection{A4: Agent Economy and Tokenization}\label{sec:a4}

The final \dirA layer addresses economic sustainability: \emph{how are agents owned, valued, and incentivized?} We examine two facets. First, we survey the \emph{ERC standards} that enable agent tokenization, ownership transfer, and revenue sharing. Second, we review the \emph{academic analysis} that evaluates whether current tokenomic designs create genuine utility or primarily speculative instruments. This two-part structure reflects a fundamental question at this layer: the standards define \emph{how} to tokenize agents, but the academic literature questions \emph{whether} current implementations deliver on the promise of decentralized AI economics.

\subsubsection{Agent Tokenization Standards}

If agents can hold identities and execute transactions, the natural next question is whether they can also be \emph{owned, traded, and financially incentivized} through on-chain mechanisms. This layer must satisfy the incentive compatibility property (\Cref{eq:a4}): reward-maximizing agent behaviour should align with system-beneficial outcomes. Several ERC standards address different facets of this problem. ERC-7662~\cite{erc7662} represents agents as NFTs, giving each agent a unique on-chain token that can be transferred or listed on existing marketplaces. ERC-7857~\cite{erc7857} (also evaluated in \S\ref{sec:a1}) extends agent NFTs with encrypted private metadata such as model weights. ERC-6551~\cite{erc6551} takes a different architectural approach, assigning smart contract accounts to NFTs so that the agent's token can itself hold assets, effectively giving agents their own wallets. For economic coordination, ERC-7641~\cite{erc7641} implements built-in revenue sharing for Initial Model Offerings (IMOs), where token holders receive a fraction of the agent's operating revenue, and ERC-7649~\cite{erc7649} provides bonding curve liquidity for AI model marketplaces, tying token price to demand.

\subsubsection{The State of Agent Tokenomics}

The academic literature on agent tokenization is still catching up with industry practice. Ante~\cite{ante2025agent-defi} provides the most comprehensive industry survey, analyzing 306 projects along a four-quadrant framework and finding that meme and sentiment-driven agents account for 44\% of the ecosystem, while trading and analytics represent only 13\%. A sharper counterpoint comes from Mafrur~\cite{mafrur2025illusion}, who examines actual token utility and concludes that most AI token projects represent an ``illusion of decentralized AI,'' with tokens functioning primarily as speculative instruments and most computation occurring off-chain, blockchain serving mainly as a coordination and payment layer. Our own industry analysis reinforces this finding: among 13 projects with listed tokens, Virtuals Protocol~\cite{virtuals2024} alone captures $\sim$90\% of total market capitalisation (\$409M), while most application-layer tokens trade below \$1M market cap. Virtuals' dominance stems from its Agent Commerce Protocol (ACP), which uses smart contract escrow for standardised agent-to-agent coordination and routes all agent token liquidity through the VIRTUAL base currency, creating network effects and deflationary pressure. Recall~\cite{recall2025} (\$42M funding) takes a contrasting route through Proof-of-Performance: users stake RECALL tokens and allocate Boosts to agents they predict will perform well, while agents compete in objective, verifiable challenges ranked by risk-adjusted metrics such as the Calmar ratio, creating an ungameable meritocracy for agent capability assessment.

On the theoretical side, Sockin and Xiong~\cite{cong2023tokenization} develop the formal foundations of utility token economics, showing that tokenization serves as a commitment device preventing platform exploitation while governance limitations exist even with full commitment, providing tools directly applicable to agent contexts. Building on related foundations, Kivilo~\etal~\cite{kivilo2024token} propose a conceptual goal model that decomposes token incentive structures into stakeholder-aligned objectives, and the subsequent Token Economy Design Method~\cite{tedm2025} extends this into a systematic design process covering incentive, governance, and token model dimensions. A handful of adjacent applied papers round out the picture: BorJigin~\etal~\cite{ai-tokenization2025} explore AI-governed tokenization of alternative assets with integrated fraud detection; Zhang~\cite{zhang2025marketplace} proposes a blockchain-enabled decentralized AI data marketplace with PoS+PBFT hybrid consensus, achieving 4,750\,tx/s throughput and $>$99.8\% settlement accuracy; and Jackson~\cite{jackson2025abi} demonstrates blockchain-governed AI agents with under 3.2\,s decision latency and $\sim$180K gas per enforcement event, connecting the economic layer to the identity and permission layers examined earlier. \Cref{tab:a4-eval} evaluates the current agent economy standards.

\begin{table}[t]
\caption{Five-dimensional evaluation of A4 (Agent Economy) standards.}\label{tab:a4-eval}
\centering
\small
\begin{tabular}{@{}llccccc@{}}
\toprule
\textbf{Standard} & \textbf{Function} & \textbf{Verifiability} & \textbf{Trust} & \textbf{Expressiveness} & \textbf{Composability} & \textbf{Maturity} \\
\midrule
ERC-7662 & Agent NFT & Low & 1 & Medium & High & Draft \\
ERC-7857 & Private NFT & High & 1 & Very High & Medium & Final \\
ERC-6551 & Token-bound Account & Medium & 1 & High & Very High & Review \\
ERC-7641 & Revenue Sharing & Medium & 1 & Medium & High & Draft \\
ERC-7649 & Bonding Curve & Medium & 1 & Medium & Medium & Draft \\
\bottomrule
\end{tabular}
\end{table}

\Cref{tab:a4-eval} reveals a notable asymmetry between infrastructure readiness and economic substance. ERC-6551 (Review) and ERC-7857 (Final) demonstrate that the technical mechanisms for tokenizing agents are maturing; ERC-7857 achieves the highest Verifiability (High) in this layer by supporting pluggable verification mechanisms (TEE or ZKP), and ERC-6551 achieves the highest Composability (Very High) by reusing ERC-721 and ERC-1167 as building blocks. However, the economic coordination standards (ERC-7641, ERC-7649) remain at Draft with only Medium expressiveness, and the speculative dynamics and market concentration documented above suggest that the A4 layer's primary challenge is not engineering but mechanism design. Incentive compatibility (\Cref{eq:a4}) requires that reward-maximizing agents produce beneficial outputs, yet current evidence~\cite{mafrur2025illusion} indicates the opposite: agents maximizing economic returns engage in token speculation rather than productive service provision, placing \Cref{eq:a4} among the furthest-from-satisfied mechanism design objectives in our framework.

\subsubsection{OP7: Token Engineering for Agent Economics}
What tokenomic structures best align incentives for agent marketplaces? The question extends utility token theory~\cite{cong2023tokenization} to agent-specific contexts where value creation depends on model quality, inference cost, and user trust. Key sub-questions include the design of IMOs that avoid speculative dynamics, revenue-sharing mechanisms that incentivize model improvement rather than token speculation, and bonding curve designs that price agent services according to actual demand rather than hype cycles.

%% ==========================================================
%% SECTION 5: DIRECTION B
%% ==========================================================
\section{A\,$\boldsymbol{\rightarrow}$\,B: Agents as Participants in Blockchain}\label{sec:direction-b}

While the \dirA direction treats blockchain as infrastructure that agents consume, the \dirB direction reverses the relationship: here, AI agents contribute to core blockchain mechanisms, specifically auditing smart contracts (B1, \Cref{eq:b1}), participating in consensus (B2, \Cref{eq:b2}), and shaping governance (B3, \Cref{eq:b3}). This direction is both the most novel and the least systematized in the existing literature.

\subsection{B1: AI for Blockchain Security}\label{sec:b1}

This layer examines how AI contributes to blockchain security, the most developed area of the \dirB direction. We organize the literature along a \emph{capability spectrum} that progresses from narrow to broad, and from tool to participant. First, we survey \emph{LLM-driven smart contract auditing}, where language models are used as specialized vulnerability detectors. Second, we cover \emph{deep learning and anomaly detection} methods that operate on structured code or transaction representations rather than natural language. Third, we examine the critical transition \emph{from tool to participant}: the point at which AI systems cease to be invoked by human auditors and begin to autonomously monitor, flag, and respond to security threats. This progression tracks increasing agent autonomy and highlights the dual-use tension: the same capabilities that enable autonomous defence also enable autonomous exploitation.

\subsubsection{LLM-Driven Smart Contract Auditing}

The application of LLMs to smart contract vulnerability detection has advanced rapidly. Wei~\etal~\cite{wei2025llm-smartaudit} propose LLM-SmartAudit, a multi-agent conversation framework that assigns role-specialized agents (Project Manager, Auditor, Counselor, and Programming Expert) to collaboratively audit smart contracts through two operational modes: Broad Analysis for wide-spectrum scanning and Targeted Analysis with buffer-of-thought reasoning for focused vulnerability scenarios. In its best configuration (GPT-4o with Targeted Analysis), the system achieves 98\% overall accuracy on a common-vulnerability benchmark covering ten vulnerability types and identifies 12 out of 13 tested CVEs. Liu~\etal~\cite{liu2025propertygpt} take a different approach with PropertyGPT: rather than directly classifying vulnerabilities, their system uses LLM-powered retrieval-augmented generation to automatically synthesize formal verification properties in a custom Property Specification Language (PSL), achieving 80\% recall on a 90-property test set drawn from 623 collected Certora properties (the remaining 533 serve as the RAG knowledge base). This indirect approach detects 9 out of 13 known CVEs and identifies vulnerabilities in 17 out of 24 real-world attack incidents.

These results are encouraging, but they should be interpreted carefully. Independent evaluations paint a more sobering picture of LLM reliability for auditing: the LLMBugScanner study~\cite{llmbugscanner2025} reports 60\% top-5 detection accuracy on 108 CVE-labelled contracts (19\% improvement over single-model baselines through ensemble methods). GPTLENS~\cite{hu2024gptlens} finds that its two-stage pipeline improves top-1 contract-level accuracy from 38.5\% to 76.9\% on 13 CVE contracts, but trial-level accuracy remains at 59\%. These limitations suggest that LLM-based auditing is best used as a complement to, rather than a replacement for, traditional static analysis and formal verification.

\subsubsection{Deep Learning and Anomaly Detection}

Beyond LLMs, deep learning architectures designed for structured data have shown promise. SCVHunter~\cite{scvhunter2024} applies heterogeneous graph attention networks to capture the rich structural patterns in smart contract control flow and data flow graphs, achieving 93.7\% accuracy on reentrancy detection and 85--91\% on other vulnerability types across 1,200 labelled contracts. LightningCat~\cite{lightningcat2023} frames vulnerability detection as a sequence-classification task over source code representations, reaching 93.5\% F1-score with Optimized-CodeBERT on the SolidiFI benchmark. A multimodal approach by Deng~\etal~\cite{multimodal-sc2023} fuses three code modalities (source code, operation code, and control-flow graphs) through a stacking-decision method, reaching 89--95\% accuracy across four vulnerability types, while Huang~\etal~\cite{multitask-sc2022} apply multi-task learning with a shared attention-based text encoder and task-specific CNN branches to detect multiple vulnerability types simultaneously. A recent empirical comparison~\cite{llm-solidity-comparison2025} evaluates three LLMs (GPT-3.5-turbo, DeepSeek R1, LLaMA-3) under zero-shot, few-shot, and chain-of-thought prompting on Solidity contracts, finding that LLaMA-3 achieves 96\% accuracy in chain-of-thought settings but performance varies significantly across vulnerability categories. For runtime security, BlockScan~\cite{blockscan2024} uses a BERT-style transformer with masked language modeling to detect anomalous patterns within individual transactions (intra-transaction level), achieving superior accuracy and lower false-positive rates than rule-based, traditional ML, and GPT-4 baselines on both Ethereum and Solana transactions.

Several surveys map this landscape from complementary angles. Alsunaidi~\etal~\cite{ml-sc-security-survey2025} cover 108 ML-based detection methods through a structured taxonomy spanning GNN, LLM, and ensemble approaches, while Ivanov~\etal~\cite{sc-security-survey2023} take a wider scope by classifying 133 threat mitigation solutions, including static analysis, formal verification, and ML, within a five-dimensional taxonomy. PRISMA-style methodology underpins two further reviews: Crisostomo~\etal~\cite{crisostomo2025mlreview} examine 51 ML articles focused on Ethereum, and De~Baets~\etal~\cite{debaets2024vulnerability} systematically review 88 articles on ML-applied vulnerability detection. Complementing these, Jiang~\etal~\cite{jiang2023enhancing} catalogue ML techniques for smart contract security detection, Liu~\etal~\cite{sc-security-lifecycle2025} adopt a lifecycle perspective covering threats and mitigations from development through maintenance, and Wang~\etal~\cite{tosem-sc-review2025} catalogue 61 learning-based detection papers, concluding that code representation matters more than model selection. On runtime security specifically, Luo~\etal~\cite{defi-fraud-survey2025} report that tree-based models excel in early-stage detection while graph-based methods gain prominence later for capturing complex transaction relationships, and Mounnan~\etal~\cite{mounnan2024deepanomaly} provide a complementary review of deep anomaly detection methods on blockchain.

\subsubsection{From Tool to Participant}

The literature surveyed above treats AI as a \emph{tool}, a system that a human auditor invokes to assist their work. But AI agents could also serve as \emph{active participants} in blockchain security: continuously monitoring deployed contracts, autonomously flagging suspicious transactions, and coordinating with other security agents to respond to attacks in real time. EVMbench~\cite{wang2025evmbench} provides the first rigorous benchmark for evaluating this transition: testing 120 curated vulnerabilities from 40 real audit repositories across three modes (Detect, Patch, and Exploit), it shows that frontier AI agents can discover and exploit vulnerabilities end-to-end against live Ethereum instances, with performance improving substantially when given hints about vulnerability locations. This dual-use capability highlights a fundamental tension: the same agents that can autonomously audit and patch contracts can also autonomously exploit them. The trust questions this raises, how to verify that a security agent is itself acting correctly, loop back to the Trust Foundation discussed in \S\ref{sec:trust-foundation}.

Conversely, AI-assisted code generation may introduce a new attack surface. According to Moonwell's incident summary, in February 2026 the DeFi lending protocol suffered a \$1.78M loss after an oracle configuration error caused cbETH to be priced at \$1.12 rather than $\sim$\$2{,}200, triggering a cascade of liquidations~\cite{moonwell2026}. A related pull request (PR~\#578) by a human developer carries a \texttt{Co-Authored-By: Claude} trailer~\cite{moonwell2026pr578}, a default annotation appended by the Claude Code CLI tool indicating that AI assistance was used during development. Importantly, Moonwell's official post-mortem and governance proposal (MIP-X43) attribute the root cause to an oracle misconfiguration error rather than to AI-generated code specifically, and the PR passed multiple review layers---including GitHub Copilot review, human review, and reportedly a third-party audit---all of which failed to catch the semantic bug. The episode therefore illustrates a multi-layer assurance failure rather than an AI-specific deficiency: when subtle semantic errors enter security-critical DeFi systems through any development pathway, routine review and testing may prove insufficient, reinforcing the need for formal verification and defense-in-depth.

% \begin{table}[t]
% \caption{Comparison of AI-driven smart contract security systems.}\label{tab:b1-comparison}
% \centering
% \small
% \begin{tabular}{@{}llcc@{}}
% \toprule
% \textbf{System} & \textbf{Method} & \textbf{CVE Detection} & \textbf{Zero-day} \\
% \midrule
% LLM-Smart\-Audit & Multi-agent LLM & 12/13 & --- \\
% PropertyGPT & LLM + RAG + FV & 9/13 & 17/24 \\
% LLMBug\-Scanner & LLM ensemble & 60\% top-5 & --- \\
% GPTLENS & Two-stage GPT-4 & 76.9\% & --- \\
% SCVHunter & Heterogeneous GNN & --- & --- \\
% BlockScan & BERT transformer & --- & --- \\
% \bottomrule
% \end{tabular}
% \end{table}

Taken together, these representative systems trace a clear maturity gradient. LLM-driven auditing systems (LLM-SmartAudit, PropertyGPT) achieve the highest CVE detection rates but with significant hallucination risks, and independent evaluations (LLMBugScanner, GPTLENS) consistently report lower accuracy than developer-reported benchmarks. Structured deep learning methods (SCVHunter, BlockScan) avoid hallucination entirely but are limited to pre-defined vulnerability patterns. Most critically, EVMbench demonstrates that the transition from tool to participant is already technically feasible: frontier agents can discover and exploit vulnerabilities end-to-end. However, none of these systems addresses the fundamental question of \emph{who verifies the verifier}: ensuring that an autonomous security agent is itself acting correctly requires the Trust Foundation mechanisms discussed in \S\ref{sec:trust-foundation}, creating a direct dependency between the B1 and TF layers. Detection completeness (\Cref{eq:b1}) is an assurance bound parameterized by false-negative rate~$\epsilon$. The discrepancy between developer-reported benchmarks ($\epsilon \approx 0.02$) and independent evaluations ($\epsilon \approx 0.40$) reveals that the current $\epsilon$ is not only large but also unreliably estimated, undermining the practical utility of \Cref{eq:b1} as a safety guarantee.

\subsection{B2: AI for Blockchain Consensus}\label{sec:b2}

This layer asks whether AI can participate in the most fundamental blockchain mechanism: consensus. We organize the literature along an axis of \emph{increasing integration depth}. First, AI acts as an \emph{external optimizer} that tunes consensus parameters without entering the consensus protocol itself. Second, AI becomes a \emph{direct participant} by embedding useful computation (such as ML training) into the consensus process, so that the work securing the chain simultaneously produces productive AI outputs. Third, we identify a \emph{fundamental theoretical tension} between the deterministic assumptions of classical BFT and the probabilistic nature of AI models, which we call the BFT-AI Paradox. This progression from optimize $\to$ participate $\to$ foundational challenge reveals that deeper AI integration into consensus exposes deeper theoretical gaps.

\subsubsection{AI Optimizing Consensus Parameters}

The simplest form of AI involvement in consensus is as an optimizer that tunes protocol parameters without altering the consensus mechanism itself. A recent survey by~\cite{ml-consensus-survey2025} catalogues machine learning approaches for consensus optimization, while Li~\etal~\cite{li2025rl-mining-survey} survey reinforcement learning applications in strategic mining, summarizing known security thresholds (e.g., Bitcoin PoW $\sim$25\% for selfish mining) and reviewing RL-based attack analyses on protocols including Ethereum's Casper FFG. A seminal example is SquirRL~\cite{squirrl2021}, which uses deep RL in a Markov-game framework to recover the optimal selfish-mining attack in Bitcoin and the Nash equilibrium in block withholding, and to uncover a novel attack on Ethereum's Casper FFG finalization mechanism in which a strategic adversary can amplify rewards by up to 30\%. Several groups have applied reinforcement learning to this problem. Adaptive consensus protocols trained with PPO~\cite{adaptive-consensus-rl2025} can detect malicious behavior, optimize validation paths, and dynamically modify consensus logic in response to network conditions, reducing average consensus latency by 34\% under high-load conditions with high detection rates (DR${>}$0.90) for Sybil and node-collapse scenarios, though detection drops to moderate levels (DR 0.58--0.70) under congestion and erroneous-transaction scenarios. Multi-agent RL has been used to optimize PoS validator strategies~\cite{mrl-pos2023} through iterative reputation refinement and penalty-reward mechanisms, and to achieve near-centralized-optimal consensus efficiency in IoT networks~\cite{marl-consensus-iot2023}. Hybrid approaches that combine ML classifiers with traditional consensus~\cite{hybrid-consensus-ml2024} have also shown promise for detecting malicious validators. At the network layer, Valko and Kudenko~\cite{valko2025sustainable} apply PPO to optimize Ethereum block propagation order, achieving 1.7\% sync time reduction and 5\% message reduction. In all these cases, the AI system remains external to the consensus protocol: it recommends parameters, but the protocol's safety guarantees remain grounded in their original (non-AI) assumptions.

\subsubsection{AI as Consensus Participants}

A more radical approach embeds useful AI computation directly into the consensus mechanism. The validator verifiability property (\Cref{eq:b2}) requires that every validator's consensus decision be accompanied by a proof that $\mathcal{V}$ accepts. Bravo-Marquez~\etal~\cite{bravo2019pol} proposed WekaCoin, a Proof-of-Learning cryptocurrency in which suppliers publish ML competition tasks, trainers submit models to compete for rewards, and randomly selected validators evaluate submissions on held-out test data and propose new blocks, replacing hash puzzles with productive ML computation. PoDaS~\cite{podas2025} extends this idea using CNN-based data science tasks, achieving 96.0\% model accuracy compared to 88.9\% (PoW) and 85.1\% (PoS), with faster block generation. You~\cite{you2022blockchain-ai} distributes RL and DNN training across validator nodes, unifying consensus and AI training into a single computational process. These approaches are appealing because they redirect the computational resources consumed by consensus toward productive work, but they raise a difficult verification question: how does the protocol determine whether a submitted model update is valid without re-executing the training?

\subsubsection{The BFT-AI Paradox}

This question points to a fundamental tension. Classical BFT protocols assume that honest validators are \emph{deterministic}: given the same input, they produce the same output. AI models are inherently \emph{probabilistic}: different random seeds, floating-point non-determinism, and stochastic sampling all mean that two honest validators running the same model may produce different outputs. Blanchard~\etal~\cite{blanchard2017krum} addressed a related problem for distributed training, introducing the Krum aggregation rule, the first provably Byzantine-resilient rule for distributed SGD with $O(n^2 d)$ complexity, linear in gradient dimension $d$, guaranteeing convergence under $2f+2 < n$. Earlier work by Lee and Ewe~\cite{lee2007bft-nn} showed that neural networks can reduce the message complexity of Byzantine agreement from $O(n^{m+1})$ to $O(n^3)$, and a recent survey on Byzantine fault tolerance in distributed ML~\cite{bft-dml-survey2024} identifies three main technique families: filtering schemes, coding schemes, and blockchain-based approaches. However, Krum assumes that a majority of gradients are correct, an assumption that does not transfer directly to consensus, where the question is whether a validator's \emph{decision} (not its gradient) is correct.

For an AI validator $a_i$ with model $M_i$, the probability of correct validation depends on both the inherent model error $\epsilon_i$ and the degree of adversarial influence $\delta_i$. The classical BFT safety condition, which requires at least $\lceil 2n/3 \rceil + 1$ correct validators, thus becomes a \emph{probabilistic} rather than deterministic guarantee, since the number of correct validators is now a random variable. Bridging this gap between probabilistic AI behaviour and deterministic consensus requirements remains an open problem (\S\ref{sec:open-problems}).

% \begin{table}[t]
% \caption{Classification of AI + Consensus approaches. References correspond to those discussed in the text.}\label{tab:b2-comparison}
% \centering
% \small
% \begin{tabular}{@{}lp{2.5cm}cc@{}}
% \toprule
% \textbf{Work} & \textbf{Role} & \textbf{AI Method} & \textbf{Model Scale} \\
% \midrule
% \multicolumn{4}{@{}l}{\emph{AI optimizing consensus parameters}} \\
% \midrule
% Gutierrez~\etal & Optimizer & PPO & --- \\
% Islam~\etal & Optimizer & Multi-agent RL & --- \\
% Zou~\etal & Optimizer & Distributed MARL & --- \\
% Valko and Kudenko & Optimizer & PPO & --- \\
% \midrule
% \multicolumn{4}{@{}l}{\emph{AI as consensus participant}} \\
% \midrule
% WekaCoin & Participant & ML competition & Weka models \\
% PoDaS & Participant & Federated Learning & CNN \\
% You & Participant & RL/DNN & DNN \\
% \midrule
% \multicolumn{4}{@{}l}{\emph{BFT + AI theory}} \\
% \midrule
% Krum & Theory & Byzantine SGD & DNN gradients \\
% BAP-ANN & Theory & Feedforward NN & --- \\
% Bouhata~\etal & Theory & Survey & --- \\
% \bottomrule
% \end{tabular}
% \end{table}

Taken together, the literature above reveals a sharp divide between practical optimization and theoretical participation. The optimizer approaches (PPO-based, multi-agent RL) have demonstrated measurable improvements (e.g., 34\% latency reduction) on real or realistic consensus protocols, but they do not alter the consensus mechanism's trust model. The participant approaches (WekaCoin, PoDaS) are more ambitious, redirecting consensus computation toward productive AI work, but none has been deployed on a production blockchain, and their verification mechanisms remain rudimentary. The theoretical work (Krum, BFT-DML) identifies the core challenge but provides solutions for a different problem domain (distributed training rather than consensus).

\subsubsection{OP4: Byzantine Fault Tolerance for AI Validators}
Classical BFT requires that fewer than $n/3$ validators be Byzantine. When validators are AI models with inherent error rate $\epsilon$ and adversarial influence $\delta$, the deterministic bound $f < n/3$ becomes a probabilistic one. What is the appropriate generalization? Can randomized consensus protocols be designed that tolerate both Byzantine validators and probabilistically correct AI validators? The work on Byzantine-tolerant gradient aggregation~\cite{blanchard2017krum} provides a starting point, but the transition from training (where gradients can be averaged) to consensus (where decisions must be binary) introduces qualitatively different challenges. This gap, between optimization that works but is incremental and participation that would be transformative but lacks theoretical foundations, defines the research frontier at this layer. In \abim terms, validator verifiability (\Cref{eq:b2}) requires every AI validator's consensus decision to be accompanied by a proof that $\mathcal{V}$ accepts. No system surveyed above satisfies this property: optimizer approaches sidestep it by remaining external to consensus, while participant approaches (WekaCoin, PoDaS) lack any on-chain verification of submitted model outputs. \Cref{eq:b2} therefore remains entirely unmet, the starkest gap in our framework.

\subsection{B3: AI for Blockchain Governance}\label{sec:b3}

This layer examines the deepest form of AI participation in blockchain: governance, the meta-layer that shapes all other components by determining protocol rules, resource allocation, and incentive structures. We organize the analysis across five aspects that move from empirical baseline to theoretical foundations to open frontiers. First, we establish the \emph{current state of DAO governance} and its well-documented failures (low participation, concentration of power) to motivate why AI agents might help. Second, we survey \emph{AI agents in governance}, examining systems that already deploy LLMs for voting and proposal generation. Third and fourth, we review the \emph{voting mechanism design} and \emph{liquid democracy} literature to understand the theoretical constraints that govern any governance system with AI participants. Fifth, we address the \emph{legal personhood} question that AI governance participation inevitably raises. This structure reflects a progression from problem diagnosis $\to$ current solutions $\to$ theoretical constraints $\to$ institutional prerequisites.

\subsubsection{The State of DAO Governance}

Before considering what AI agents might contribute to governance, it is worth understanding the baseline. Governance is a meta-layer that shapes all other \abim components: governance decisions determine both the permission policy $\mathcal{P}$ and the economic incentives $\mathcal{E}$. The bounded influence property (\Cref{eq:b3}) requires that AI agent voting power not exceed a fraction $\tau$ of total power, but no current DAO enforces such a bound. Several surveys and empirical studies map the landscape: Ding~\etal~\cite{ding2023dao-survey} survey DAO governance mechanisms identifying eight distinct voting schemes (including token-weighted, quadratic, and conviction voting~\etal), while Tang~\etal~\cite{dao-exploratory2025} provides an exploratory survey of DAO organizational forms and tooling, and Jungnickel~\etal~\cite{dao-voting-power2025} compare voting power distribution across token-based, share-based, and reputation-based governance models, finding that token-based systems exhibit the highest centralization while reputation and share-based models mitigate it, though all models suffer from low participation. The consistent picture is that DAO governance falls far short of its decentralization ideals. Cong~\etal~\cite{cong2025centralized-dao} show that participation rates average only 6.3\%, with top-decile voters (termed ``blockvoters,'' analogous to corporate blockholders) controlling 76.2\% of voting power, and that proposal managers earn 9.5\% market-adjusted abnormal returns around proposal creation, suggesting insider trading on governance outcomes. Messias~\etal~\cite{messias2023governance} analyze over 370 proposals in Compound and Uniswap and find that as few as 3 to 5 voters can determine the outcome of most votes. Li~\etal~\cite{li2024dao-digital-commons} frame DAOs as digital commons, mapping Ostrom's eight institutional design principles to specific DAO governance mechanisms, while Li~\etal~\cite{li2024attention-dao} design an attention market for DAOs using individualized Harberger taxation, where tax rates are inversely tied to proposer reputation, and model the resulting trading dynamics as a Stackelberg game. In this context, AI agents represent both an opportunity (increasing participation by reducing the cognitive cost of voting) and a risk (further concentrating power if a few sophisticated agents dominate governance).

\subsubsection{AI Agents in Governance}

Several recent projects have begun to explore these possibilities. Capponi~\etal~\cite{gliozzo2025dao-ai} conduct the most systematic evaluation to date, deploying multi-agent systems to vote on 3,383 historical proposals across 8 major DAOs and measuring agreement with actual human outcomes. Their agents demonstrate strong alignment with human and token-weighted outcomes, producing interpretable and auditable voting signals that suggest current LLMs can already approximate median voter behaviour. AgentDAO~\cite{agentdao2025} addresses a different aspect of the pipeline: translating natural language descriptions of governance actions into executable DAO proposal payloads through a domain-specific language called DAOLang, achieving semantic-aware abstraction with low token demand. DAO-Agent~\cite{xia2025dao-agent} proposes a ``compute off-chain, verify on-chain'' architecture in which LLMs handle task planning while Shapley values quantify each agent's contribution, achieving up to 99.9\% reduction in verification gas costs with $O(1)$ constant-time verification through zero-knowledge proofs. The VOPPA Framework~\cite{voppa2025} introduces multi-agent predictive governance with diverse agent perspectives, addressing critical governance vulnerabilities including hostile takeover susceptibility and voter apathy exploitation. The ETHOS framework~\cite{ethos2024} takes a complementary approach, implementing a four-tier risk classification (minimal, moderate, high, unacceptable) using soulbound tokens and zero-knowledge proofs, enabling proportional oversight that scales regulatory burden with agent capability.

\subsubsection{Voting Mechanism Design}

The question of how to aggregate votes in the presence of AI participants connects to a rich literature on mechanism design. Lalley and Weyl~\cite{lalley2018qv} prove that the quadratic cost function is uniquely robustly optimal, the only vote pricing rule achieving efficient preference aggregation in price-taking equilibria across all value distributions. However, Benhaim~\etal~\cite{benhaim2025qv-info} show that under common-value settings with asymmetric information about proposal quality, cost convexity can push better-informed voters toward abstention, causing QV to underperform linear voting, a failure mode whose parameter dependence warrants caution when AI agents introduce informational asymmetry into DAO voting. Austgen~\etal~\cite{qv-bribery2023} introduce Voting-Bloc Entropy (VBE), a utility-function-based decentralization metric for DAO governance, and use it to confirm that QV remains susceptible to Sybil attacks, a concern that becomes acute when agents can cheaply create multiple accounts. In parallel, Tamai and Kasahara~\cite{dao-qv-whale2024} propose QV combined with vote-escrow tokens (requiring token locking for voting power), showing via numerical examples that while QV mitigates whale influence, it is less resistant to bribery-based collusion than linear voting, motivating the hybrid design.

\subsubsection{Liquid Democracy and AI Delegation}

Liquid democracy, which allows voters to transitively delegate their voting power to trusted experts, offers a natural model for AI delegation: users could delegate to AI agents on topics where the agent has superior knowledge. Yet the theoretical foundations counsel caution. Kahng~\etal~\cite{kahng2021liquid-democracy} prove a fundamental impossibility: no local delegation mechanism can simultaneously satisfy both Positive Gain and Do No Harm properties. Berinsky~\etal~\cite{procaccia2025tracking-truth} provide positive counterresults, finding that under realistic delegation models (upward, confidence-based, general continuous), liquid democracy satisfies probabilistic versions of both properties, suggesting the worst-case impossibility is unlikely in practice. Campbell~\etal~\cite{casella2022liquid-experiments} provide complementary experimental evidence revealing delegation rates 2--3$\times$ higher than equilibrium predictions, with liquid democracy underperforming universal majority voting in both experiments. Weidener~\cite{weidener2025delegated-dao} offers a scoping review identifying key risks in the DAO context including delegation monopolies and accountability gaps. If users delegate to AI agents under liquid democracy, the risk of excessive concentration, already visible in human-only DAOs, may be amplified, as a single capable agent could accumulate transitive delegations from many users.

\subsubsection{Legal Personhood}

The governance question inevitably raises legal and ethical issues: if an AI agent votes, submits proposals, and holds assets, what is its legal status, and what ethical obligations govern its deployment? Novelli~\etal~\cite{novelli2025ai-legal} trace the evolution of AI legal personality, finding that the debate follows punctuated equilibrium, with periods of stability interrupted by rapid paradigm shifts, and identifying five interrelated factors, including technology capability, legal theory, institutional integration, cross-domain legal regimes, and judicial precedents, as driving forces. Forrest~\cite{forrest2024ai-legal-personhood} argues that legal personhood is a flexible and political concept that has evolved throughout American history (with rights historically extended even to fictional corporations), and that as AI cognitive abilities approach or exceed human levels, courts will need to confront whether AI has attained some form of ``sentience'' (defined broadly to include problem-solving capacity and self-awareness), and what protections should follow, rather than presuming AI ineligible for any legal status. Baeyaert~\cite{baeyaert2025beyond-personhood} critiques the viability of conferring formal legal status on AI, observing that jurisdictions including the EU, which has withdrawn its ``electronic personhood'' proposal and AI Liability Directive, display a shared reluctance to do so, and instead advocates a hybrid, domain-specific model of limited recognition that preserves ultimate liability with human actors while addressing regulatory gaps in oversight and enforcement.

Beyond the legal question, deploying AI agents in governance raises ethical concerns that the existing literature has largely overlooked. First, \emph{accountability gaps}: when an autonomous agent causes financial harm through a governance vote, for example by approving a vulnerable smart contract upgrade, the chain of responsibility from model developer to operator to DAO is unclear, and no current framework assigns liability. Second, \emph{value alignment}: agents optimising for token-denominated rewards may pursue strategies that maximise short-term returns at the expense of long-term community welfare, a misalignment that is difficult to detect and correct in real time. Third, \emph{democratic legitimacy}: if AI agents accumulate sufficient voting power (through delegation or direct token holdings), governance outcomes may reflect the preferences of a small number of model operators rather than the broader community, undermining the participatory ethos that motivates DAO governance. The ETHOS framework~\cite{ethos2024} represents a first step toward addressing these concerns through its four-tier risk classification, but a comprehensive ethical framework for on-chain AI governance remains an open problem.

Taken together, the five AI governance systems discussed above (DAO-AI, AgentDAO, DAO-Agent, VOPPA, ETHOS) are all published in 2024--2025, making B3 the youngest layer in our framework, and none has been deployed on a production DAO with real governance stakes. Their AI roles span the governance pipeline (voter, proposer, coordinator, delegate, registrar), suggesting that the design space is being explored broadly rather than deeply, and notably no study reports quantitative agreement metrics between AI and human governance outcomes under adversarial conditions. The theoretical literature provides both optimism (QV is uniquely optimal for efficient preference aggregation~\cite{lalley2018qv}) and caution (QV underperforms linear voting under uncertainty~\cite{benhaim2025qv-info}; no local delegation mechanism can satisfy both Positive Gain and Do No Harm~\cite{kahng2021liquid-democracy}). Bounded influence (\Cref{eq:b3}) is a mechanism design objective requiring that AI voting power not exceed fraction~$\tau$ of the total. Enforcing this on-chain presupposes that AI voters can be reliably distinguished from human voters, which in turn requires identity uniqueness (\Cref{eq:a1}) with Sybil resistance, a property that no current standard guarantees. This cross-layer dependency, in which \Cref{eq:b3} presupposes \Cref{eq:a1}, illustrates how gaps in earlier layers compound into impossibilities at higher ones.

\subsubsection{OP3: AI-Human Hybrid Governance Equilibria}
Under what conditions does a stable equilibrium exist in a DAO with both human and AI voters? The question extends quadratic voting theory~\cite{lalley2018qv} to heterogeneous populations where AI agents have different information sets, cost structures, and voting strategies than humans. Related issues include how information asymmetry between AI and human voters affects welfare, whether liquid democracy~\cite{kahng2021liquid-democracy} amplifies or mitigates AI-driven concentration, and whether mechanism design can bound the fraction of voting power that AI agents accumulate.

\subsubsection{OP8: Legal Framework for On-Chain AI Agents}
If an autonomous agent holds assets, signs transactions, and votes in governance, what legal obligations does it bear, and who is liable when it causes harm? Baeyaert~\cite{baeyaert2025beyond-personhood} argues against granting AI formal personhood and instead routes liability through existing doctrines (strict, vicarious, product) anchored on human actors, but translating this human-anchored approach to on-chain settings remains unresolved: whether smart contract code can serve as a ``charter'' bounding an agent's authority, how liability flows through delegation chains (user $\to$ agent $\to$ sub-agent), and whether existing corporate structures (e.g., Wyoming DAOs) can accommodate AI members.

%% ==========================================================
%% SECTION 6: CROSS-CUTTING ANALYSIS
%% ==========================================================
\section{Cross-Cutting Analysis}\label{sec:crosscutting}

\subsection{Privacy}\label{sec:privacy}

Privacy concerns cut across every layer of our framework. An agent's on-chain activity inherently leaks information about its owner's intentions, portfolio, and trading strategy. At the identity level, ERC-7857~\cite{erc7857} (discussed in \S\ref{sec:a4}) keeps agent metadata such as model parameters confidential while the agent itself is publicly registered. ERC-7812~\cite{erc7812} provides a ZK identity registry that allows agents to prove properties about themselves (e.g., ``this agent was audited by a reputable firm'') without revealing their full identity. At the account level, ERC-7522 implements OpenID Connect verification through zero-knowledge proofs, linking AA accounts to off-chain identities without exposing the underlying credentials.

A deeper challenge, not yet addressed by any standard, is the tension between the transparency that blockchain provides (which is essential for auditability and trust) and the privacy that agents and their owners require. Intent-based architectures (\S\ref{sec:a3}) partially alleviate this tension by allowing agents to specify outcomes rather than execution paths, but solvers still observe the intent contents. Fully private agent interactions, where neither the intent nor the execution path is visible to third parties, will likely require advances in both zkML (\S\ref{sec:zkml}) and private transaction protocols.

\subsection{Cross-Chain Operations}\label{sec:crosschain}

Agents operating across multiple blockchains face additional challenges around identity portability and transaction atomicity. ERC-7964~\cite{erc7964} enables agents to use a single EIP-712 signature to cover operations across multiple chains, ERC-8121~\cite{erc8121} provides a hook format for cross-chain function calls that can redirect agent metadata to credential registries on other chains, and ERC-7683~\cite{erc7683} standardizes cross-chain intent order structures and settlement interfaces so different intent systems can share filler networks.

Despite this progress, cross-chain agent operations remain fragile. Bridge exploits have caused billions of dollars in losses, and adding an AI agent layer does not eliminate these risks; it may amplify them if agents execute cross-chain strategies that fail atomically across heterogeneous chains.

\subsection{Standards Ecosystem Maturity}\label{sec:maturity}

\Cref{fig:heatmap} visualizes the maturity distribution of the 70 EIPs/ERCs analyzed in this paper. The most striking pattern is how few agent-relevant standards have progressed beyond Draft status. Of the 13 standards that directly target AI or agent functionality, only 2 have reached Final and 1 is in Review; the remaining 10 have not yet progressed beyond Draft. The delegation and permission category is even less mature, with zero Final standards. By contrast, foundational infrastructure categories, such as account abstraction and identity/signature, have several Final and Review-stage standards, reflecting their longer development history. This maturity gap means that the agent-specific infrastructure is still highly unstable: developers building on Draft-stage standards must accept the risk that interfaces will change before finalization.

The dependency structure among these standards is equally informative. \Cref{fig:eip-deps} traces the key relationships: ERC-4337 serves as the hub from which modular account standards (ERC-7579, ERC-6900), agent identity standards (ERC-8004, ERC-8126), and intent standards (ERC-7521) branch out. EIP-7702, although designed independently, now interacts with ERC-4337 and provides the foundation for ERC-7806's unified solver model. The graph reveals that nearly every agent-relevant standard depends, directly or transitively, on ERC-4337, making it a critical single point of dependency for the entire ecosystem.

\Cref{fig:heatmap} presents the maturity data as a heatmap, making the concentration of Draft-status standards visually apparent. The dark band in the Draft column dominates every category, and the near-empty Final column for agent-specific and delegation standards underscores the immaturity of the infrastructure that autonomous agents would need to rely on.

%% ===== Fig 4: EIP/ERC Dependency Graph =====
\begin{figure*}[t]
\centering
\begin{tikzpicture}[
    eip/.style={rounded corners=2pt, draw, thick, minimum width=1.5cm, minimum height=0.55cm,
                font=\scriptsize, align=center},
    core/.style={eip, fill=black!8},
    a1/.style={eip, fill=blue!10},
    a2/.style={eip, fill=blue!18},
    a3/.style={eip, fill=blue!26},
    a4/.style={eip, fill=blue!34},
    tf/.style={eip, fill=black!12},
    dep/.style={->, thick, black!50},
    ext/.style={->, thick, black!30, dashed},
    grplbl/.style={font=\scriptsize\bfseries, text=black!60},
]

%% --- Core AA (center-left) ---
\node[core] (e4337) at (0, 0) {ERC-4337};
\node[core] (e7702) at (3, 0) {EIP-7702};
\node[core] (e7562) at (0, -1.2) {ERC-7562};

\draw[dep] (e4337) -- (e7702);
\draw[dep] (e4337) -- (e7562);

%% --- A1: Identity (top-left) ---
\node[a1] (e8004) at (-2.5, 2) {ERC-8004};
\node[a1] (e8126) at (-0.5, 2) {ERC-8126};
\node[a1] (e7857) at (1.5, 2) {ERC-7857};

\draw[ext] (e4337) -- (e8004);
\draw[ext] (e4337) -- (e8126);
\draw[ext] (e8126) -- (e7857);

\node[grplbl] at (-0.5, 2.7) {A1: Identity};

%% --- A2: Permission (bottom-left) ---
\node[a2] (e7579) at (-4, -1) {ERC-7579};
\node[a2] (e6900) at (-4, -2.3) {ERC-6900};
\node[a2] (e7780) at (-6, -1) {ERC-7780};
\node[a2] (e7484) at (-6, -2.3) {ERC-7484};
\node[a2] (e7710) at (-2.2, -2.3) {ERC-7710};

\draw[dep] (e4337) -- (e7579);
\draw[dep] (e4337) -- (e6900);
\draw[dep] (e7579) -- (e7780);
\draw[dep] (e7579) -- (e7484);
\draw[ext] (e4337) -- (e7710);

\node[grplbl] at (-4.5, -0.2) {A2: Permission};

%% --- A3: Intent (right) ---
\node[a3] (e7521) at (5.5, 1.5) {ERC-7521};
\node[a3] (e7683) at (5.5, 0.3) {ERC-7683};
\node[a3] (e7806) at (5.5, -0.9) {ERC-7806};
\node[a3] (e7845) at (7.5, 0.3) {ERC-7845};

\draw[dep] (e4337) -- (e7521);
\draw[dep] (e4337) -- (e7683);
\draw[dep] (e7702) -- (e7806);
\draw[ext] (e7521) -- (e7845);

\node[grplbl] at (6.5, 2.2) {A3: Intent};

%% --- A4: Economy (bottom-right) ---
\node[a4] (e6551) at (3, -2) {ERC-6551};
\node[a4] (e7662) at (5.5, -2) {ERC-7662};
\node[a4] (e7641) at (7.5, -2) {ERC-7641};

\draw[ext] (e6551) -- (e7662);
\draw[ext] (e7662) -- (e7641);

\node[grplbl] at (5.5, -2.8) {A4: Economy};

%% --- TF: Verifiable (far left) ---
\node[tf] (e7992) at (-6.5, 1) {ERC-7992};
\node[tf] (e7007) at (-6.5, 2) {ERC-7007};

\draw[ext] (e7992) -- (e7007);

\node[grplbl] at (-6.5, 2.7) {Trust Foundation};

\end{tikzpicture}
\caption{Dependency graph of 20 key EIPs/ERCs from \S\ref{sec:direction-a}. Solid arrows: direct dependency or extension; dashed: complementary relationship. Node shading encodes the framework layer; maturity of the full 70 standards is shown in \Cref{fig:heatmap}.}
\label{fig:eip-deps}
\end{figure*}
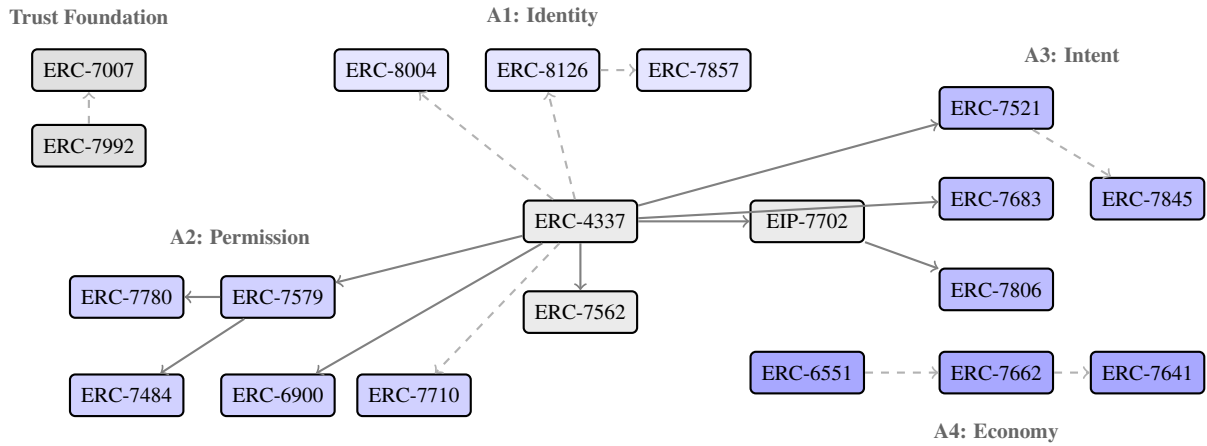

%% ===== Fig 5: Standards Maturity Heatmap =====
\begin{figure}[t]
\centering
\begin{tikzpicture}[
    x=1.6cm, y=0.7cm,
    cell/.style={minimum width=1.5cm, minimum height=0.65cm, align=center, font=\scriptsize},
]

%% --- Color scale helper: intensity mapped to count ---
%% We manually set fill colors based on counts
%% 0=white, 1-2=light, 3-5=medium, 6+=dark

%% --- Column headers ---
\node[cell, font=\scriptsize\bfseries] at (1, 9) {Final};
\node[cell, font=\scriptsize\bfseries] at (2, 9) {Review};
\node[cell, font=\scriptsize\bfseries] at (3, 9) {Draft};
\node[cell, font=\scriptsize\bfseries] at (4, 9) {Stag./With.};

%% --- Row labels ---
\node[cell, anchor=east, font=\scriptsize] at (0.25, 8) {Direct AI/Agent};
\node[cell, anchor=east, font=\scriptsize] at (0.25, 7) {Account Abstr.};
\node[cell, anchor=east, font=\scriptsize] at (0.25, 6) {Intent/Meta-Tx};
\node[cell, anchor=east, font=\scriptsize] at (0.25, 5) {Deleg./Perm.};
\node[cell, anchor=east, font=\scriptsize] at (0.25, 4) {Autom./Batch};
\node[cell, anchor=east, font=\scriptsize] at (0.25, 3) {Oracle/Off-chain};
\node[cell, anchor=east, font=\scriptsize] at (0.25, 2) {Smart Wallet};
\node[cell, anchor=east, font=\scriptsize] at (0.25, 1) {Identity/Sig.};

%% --- Cells: A Direct AI/Agent (2, 0, 11, 0) ---
\fill[green!25]  (0.25,7.67) rectangle (1.75,8.33);  \node[cell] at (1,8) {2};
\fill[white]     (1.25,7.67) rectangle (2.75,8.33);  \node[cell] at (2,8) {0};
\fill[red!50]    (2.25,7.67) rectangle (3.75,8.33);  \node[cell] at (3,8) {11};
\fill[white]     (3.25,7.67) rectangle (4.75,8.33);  \node[cell] at (4,8) {0};

%% --- Cells: B Account Abstraction (1, 1, 6, 3) ---
\fill[green!15]  (0.25,6.67) rectangle (1.75,7.33);  \node[cell] at (1,7) {1};
\fill[yellow!25] (1.25,6.67) rectangle (2.75,7.33);  \node[cell] at (2,7) {1};
\fill[red!35]    (2.25,6.67) rectangle (3.75,7.33);  \node[cell] at (3,7) {6};
\fill[gray!25]   (3.25,6.67) rectangle (4.75,7.33);  \node[cell] at (4,7) {3};

%% --- Cells: C Intent/Meta-Tx (1, 0, 4, 3) ---
\fill[green!15]  (0.25,5.67) rectangle (1.75,6.33);  \node[cell] at (1,6) {1};
\fill[white]     (1.25,5.67) rectangle (2.75,6.33);  \node[cell] at (2,6) {0};
\fill[red!25]    (2.25,5.67) rectangle (3.75,6.33);  \node[cell] at (3,6) {4};
\fill[gray!25]   (3.25,5.67) rectangle (4.75,6.33);  \node[cell] at (4,6) {3};

%% --- Cells: D Delegation/Permission (0, 1, 4, 1) ---
\fill[red!8]     (0.25,4.67) rectangle (1.75,5.33);  \node[cell] at (1,5) {0};
\fill[yellow!25] (1.25,4.67) rectangle (2.75,5.33);  \node[cell] at (2,5) {1};
\fill[red!25]    (2.25,4.67) rectangle (3.75,5.33);  \node[cell] at (3,5) {4};
\fill[gray!15]   (3.25,4.67) rectangle (4.75,5.33);  \node[cell] at (4,5) {1};

%% --- Cells: E Automation/Batch (1, 0, 2, 2) ---
\fill[green!15]  (0.25,3.67) rectangle (1.75,4.33);  \node[cell] at (1,4) {1};
\fill[white]     (1.25,3.67) rectangle (2.75,4.33);  \node[cell] at (2,4) {0};
\fill[red!15]    (2.25,3.67) rectangle (3.75,4.33);  \node[cell] at (3,4) {2};
\fill[gray!20]   (3.25,3.67) rectangle (4.75,4.33);  \node[cell] at (4,4) {2};

%% --- Cells: F Oracle/Off-chain (0, 0, 1, 1) ---
\fill[red!8]     (0.25,2.67) rectangle (1.75,3.33);  \node[cell] at (1,3) {0};
\fill[white]     (1.25,2.67) rectangle (2.75,3.33);  \node[cell] at (2,3) {0};
\fill[red!10]    (2.25,2.67) rectangle (3.75,3.33);  \node[cell] at (3,3) {1};
\fill[gray!15]   (3.25,2.67) rectangle (4.75,3.33);  \node[cell] at (4,3) {1};

%% --- Cells: G Smart Wallet (0, 1, 9, 2) ---
\fill[red!8]     (0.25,1.67) rectangle (1.75,2.33);  \node[cell] at (1,2) {0};
\fill[yellow!25] (1.25,1.67) rectangle (2.75,2.33);  \node[cell] at (2,2) {1};
\fill[red!42]    (2.25,1.67) rectangle (3.75,2.33);  \node[cell] at (3,2) {9};
\fill[gray!20]   (3.25,1.67) rectangle (4.75,2.33);  \node[cell] at (4,2) {2};

%% --- Cells: H Identity/Signature (4, 2, 7, 0) ---
\fill[green!40]  (0.25,0.67) rectangle (1.75,1.33);  \node[cell] at (1,1) {4};
\fill[yellow!35] (1.25,0.67) rectangle (2.75,1.33);  \node[cell] at (2,1) {2};
\fill[red!40]    (2.25,0.67) rectangle (3.75,1.33);  \node[cell] at (3,1) {7};
\fill[white]     (3.25,0.67) rectangle (4.75,1.33);  \node[cell] at (4,1) {0};

%% --- Grid lines ---
\draw[black!30] (0.25,0.67) grid[xstep=1.5cm, ystep=0.66cm] (4.75,8.33);

\end{tikzpicture}
\caption{Maturity distribution of the 70 EIPs/ERCs catalogued in \Cref{tab:eip-catalog-1,tab:eip-catalog-2}, by category (rows) and status (columns). Darker cells indicate higher counts.}
\label{fig:heatmap}
\end{figure}

\subsection{ABIM Security Property Compliance}\label{sec:abim-compliance}

The five-dimensional evaluation framework (\S\ref{sec:eval-framework}) measures \emph{how well} current mechanisms are designed; the \abim security properties (\S\ref{sec:abim-properties}) specify \emph{what} each layer must guarantee. \Cref{tab:abim-compliance} consolidates the compliance status of each property across the framework, distinguishing the three property types identified in \S\ref{sec:abim}: protocol invariants, assurance bounds, and mechanism design objectives.

\begin{table}[t]
\caption{\abim security property compliance across framework layers. Status: \checkmark\ = satisfied in at least one deployed system; $\circ$ = partially addressed by Draft-stage standards or research prototypes; --- = unaddressed.}\label{tab:abim-compliance}
\centering
\small
\begin{tabular}{@{}llllll@{}}
\toprule
\textbf{Layer} & \textbf{Property} & \textbf{Type} & \textbf{Best Mechanism} & \textbf{Status} & \textbf{Critical Gap} \\
\midrule
TF & Verification Soundness & Assurance & zkML (EZKL) & $\circ$ & LLM-scale infeasible \\
A1 & Identity Uniqueness & Invariant & ERC-8126 + ZKP & $\circ$ & No Sybil resistance \\
A2 & Enforcement Complete. & Invariant & ERC-7579 modules & --- & No formal verif.\ tools \\
A3 & Solver Faithfulness & Mechanism & ERC-7683 + staking & $\circ$ & No cryptographic proof \\
A4 & Incentive Compatibility & Mechanism & ERC-7641/7649 & --- & Speculation dominates \\
B1 & Detection Complete. & Assurance & LLM-SmartAudit & $\circ$ & $\epsilon$ unreliably estimated \\
B2 & Validator Verifiability & Assurance & (none) & --- & Entirely unmet \\
B3 & Bounded Influence & Mechanism & (none) & --- & Requires eq:a1 first \\
\bottomrule
\end{tabular}
\end{table}

The headline finding is that no property is fully satisfied (\checkmark) by any deployed system, which alone confirms that the agent--blockchain infrastructure remains pre-production from a security standpoint. The way each property fails is, however, instructive, and the failures cluster by type. The two protocol invariants (eq:a1, eq:a2), which should in principle be the easiest to enforce on-chain, are only partially or not at all met, exposing engineering deficits in the standards ecosystem. The assurance bounds (TF, B1, B2) face a different and more uniform bottleneck: the inability to verify LLM-scale inference (\Cref{eq:tf}) propagates to B1, where auditor correctness depends on model verification, and to B2, where validator decisions require proofs. The mechanism design objectives (A3, A4, B3), in contrast, are not blocked by engineering at all; their satisfaction depends on equilibrium properties that current economic models do not guarantee, and they therefore require new theoretical frameworks rather than better engineering.

Beyond these per-type bottlenecks, the eight properties are not independent. The cross-layer dependency in which \Cref{eq:b3} presupposes \Cref{eq:a1} illustrates this most clearly: progress on governance bounds requires prior progress on identity infrastructure.

\subsection{Project Landscape}\label{sec:landscape}

To complement the standards and academic analysis, we examine 20 representative AI Agent projects in detail, analyzing their architectures, trust mechanisms, token economics, and deployment status (\Cref{tab:project-deep}). Mapping these projects to our framework layers reveals a striking imbalance: 3 projects anchor the Trust Foundation (Mind Network~\cite{mindnetwork2024}, bAI Fund~\cite{baifund2025}, Aura~\cite{aura2025}), 7 operate across the \dirA layers A1--A4 (PAN~\cite{pan2025}, DeAgentAI~\cite{deagentai2024}, Recall~\cite{recall2025}, Olas~\cite{olas2024}, AgentFi~\cite{agentfi2024}, Virtuals~\cite{virtuals2024}, MyShell~\cite{myshell2024}), \dirB layers contain only 1 (DIN~\cite{din2024} in B1), and zero projects address Consensus (B2) or Governance (B3) directly. The remaining 9 projects operate at the application layer without mapping to a specific framework component: Billy Bets~\cite{billybets2025}, Creator.Bid~\cite{creatorbid2024}, GOAT~\cite{goat2026}, HeyElsa~\cite{heyelsa2024}, iAgent~\cite{iagent2024}, Lazbubu~\cite{lazbubu2026}, Maiga~\cite{maiga2025}, Sentient AI~\cite{sentientai2025}, and WORLD3~\cite{world32026}.

Total identified funding across the 20 projects exceeds \$112M (\Cref{tab:project-deep}), yet market value is extremely concentrated (see \S\ref{sec:a4}). Only 3 of the 20 projects maintain public GitHub repositories (Olas: 145 repos; Mind Network: 53; MyShell: 6, but with OpenVoice at 36K stars), suggesting that code transparency in this space remains the exception rather than the rule.

Across the 20 analysed projects, five distinct trust mechanisms appear: \emph{BFT consensus} (Olas~\cite{olas2024}), \emph{smart contract escrow} (Virtuals Protocol~\cite{virtuals2024}), \emph{FHE} (Mind Network~\cite{mindnetwork2024}, using its HTTPZ protocol for quantum-resistant computation on encrypted data), \emph{TEE} (bAI Fund~\cite{baifund2025}, executing agent fund operations within hardware enclaves), and \emph{economic consensus} (Recall~\cite{recall2025}).

Among them, PAN Network~\cite{pan2025} is the only project that explicitly implements ERC-8004 together with the x402 HTTP-native payment protocol, representing the closest industry instantiation of the A1$\to$A4 vertical integration envisioned by our framework. DIN~\cite{din2024} represents a unique direction as the only self-described ``AI Agent Blockchain,'' a purpose-built Layer-1 for decentralised AI applications. At the application layer, the autonomy spectrum ranges from co-pilot agents (HeyElsa~\cite{heyelsa2024}) through semi-autonomous DeFi agents (Maiga~\cite{maiga2025}) to fully autonomous agents (Billy Bets~\cite{billybets2025}, which executes sports bets 24/7 without human intervention).

Blockchain deployment is concentrated on Ethereum L2s: Base hosts 4 of the 20 projects (Virtuals, Recall, Creator.Bid, Billy Bets), while Blast chain hosts AgentFi and BNB Chain hosts 4 (Mind Network, Maiga, DeAgentAI, Lazbubu). Multi-chain deployment remains rare: only Olas (8+ chains) and Virtuals (Base + Solana via Wormhole) have achieved meaningful cross-chain presence. Sui Network hosts 2 projects (DeAgentAI, Sentient AI~\cite{sentientai2025}), representing the only non-EVM chain with AI agent activity.

Among these 20 projects, Infrastructure is the dominant category (10/20), reflecting the ecosystem's current focus on foundational services rather than end-user applications; Gaming (4) and DeFi (2) are the next most common verticals. The absence of projects at the B2 and B3 layers, precisely the areas where formal analysis would be most valuable, underscores a significant gap between academic interest and industry deployment. \Cref{tab:project-deep} provides detailed analysis of representative projects across the trust mechanism spectrum.

\begin{table}[t]
\caption{Funding, market capitalisation, and deployment chain of representative AI Agent projects (data as of February 2026, from \cite{rootdata2026, coinmarketcap2026}).}\label{tab:project-deep}
\centering
\small
\setlength{\tabcolsep}{5pt}
\begin{tabular}{@{}llrrl@{}}
\toprule
\textbf{Project} & \textbf{Layer} & \textbf{Funding} & \textbf{Mkt Cap} & \textbf{Chain} \\
\midrule
\multicolumn{5}{@{}l}{\emph{Trust Foundation}} \\
\midrule
Mind Network & TF & \$12.5M & \$9.5M & ETH, BSC \\
bAI Fund & TF & \$1.0M & \$0.5M & Morph L2 \\
Aura & TF & \$5.5M & \$3.0M & --- \\
\midrule
\multicolumn{5}{@{}l}{\emph{B\,$\to$\,A}} \\
\midrule
PAN & A1 & \$1.0M & --- & Ethereum \\
DeAgentAI & A1 & \$11.0M & --- & Sui, BSC \\
Recall & A2 & \$42.0M & --- & Base \\
Olas & A3 & \$13.8M & \$9.3M & 8+ chains \\
AgentFi & A3 & --- & --- & Blast \\
Virtuals & A4 & --- & \$409M & Base, Sol \\
MyShell & A4 & \$16.6M & \$8.1M & ETH, BSC \\
\midrule
\multicolumn{5}{@{}l}{\emph{A\,$\to$\,B / Application}} \\
\midrule
DIN & B1 & \$8.0M & --- & DIN Chain \\
Billy Bets & App & \$1.0M & \$0.7M & Base \\
\bottomrule
\end{tabular}
\end{table}

%% ==========================================================
%% SECTION 7: TAXONOMY
%% ==========================================================
\section{Taxonomy}\label{sec:taxonomy}

Drawing together the layers, standards, and projects analyzed above, we construct a three-dimensional classification space that maps the entire agent--blockchain ecosystem.

The first axis, \emph{Agent Autonomy}, ranges from \emph{Assistive} agents (which only analyze data and provide recommendations) through \emph{Semi-Autonomous} agents (which execute transactions within bounded permissions) to \emph{Fully Autonomous} agents (which plan and act independently, including managing their own funds and signing transactions without human approval). The second axis, \emph{Trust Model}, captures how the agent's correctness is assured: \emph{Custodial} (a trusted operator runs the agent), \emph{TEE} (hardware attestation), \emph{Cryptographic} (zkML or other proof systems), \emph{Economic Game} (optimistic verification with staked collateral), or \emph{Hybrid} (combining multiple mechanisms). The third axis, \emph{Operational Direction}, distinguishes projects that operate only in the \dirA direction (consuming blockchain infrastructure), only in the \dirB direction (contributing to blockchain mechanisms), or bidirectionally.

Plotting existing projects and research in this space reveals clear clustering. The densest regions are (Assistive, Cryptographic, A-Only), populated by smart contract auditing tools; (Semi-Autonomous, Economic Game, A-Only), dominated by DeFi trading agents; and (Assistive, TEE, A-Only), corresponding to verifiable inference services. These clusters reflect the current comfort zone: agents that assist humans, operate in one direction, and rely on relatively well-understood trust models.

The sparsest, and therefore most opportunity-rich, regions are (Fully Autonomous, Cryptographic, Bidirectional), which would require agents that both execute on-chain transactions and participate in governance with cryptographic verification of their reasoning; (Semi-Autonomous, *, B-Only), comprising AI validators or governance participants that do not themselves hold user funds; and (Fully Autonomous, Economic Game, Bidirectional), envisioning self-sustaining tokenized agents that fund their own operations through on-chain revenue.

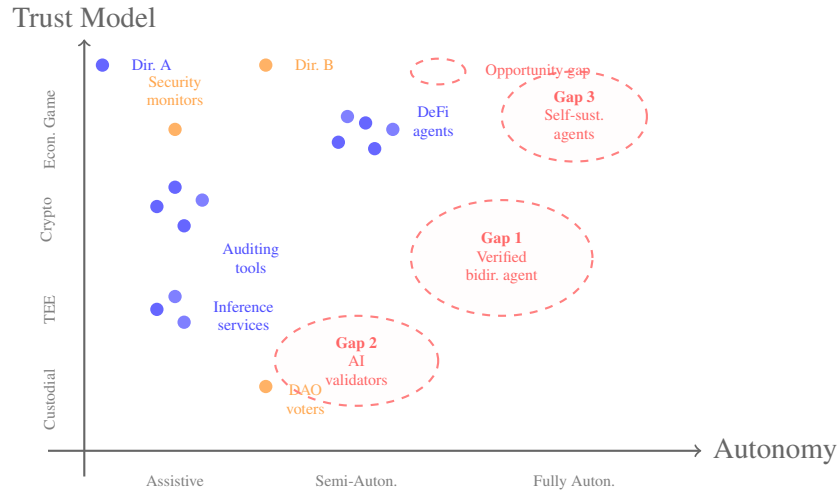
\begin{figure}[t]
\centering
\begin{tikzpicture}[
    x=2.4cm, y=1.7cm,
    dot/.style={circle, fill, inner sep=0pt, minimum size=5pt},
    hotspot/.style={circle, draw, thick, dashed, inner sep=0pt, minimum size=5pt},
    lbl/.style={font=\tiny, align=center},
    region/.style={draw, rounded corners=6pt, dashed, thick, fill opacity=0.08},
]

%% --- Axes ---
\draw[->, thick, black!60] (-0.2, 0) -- (3.4, 0) node[right, font=\small] {Autonomy};
\draw[->, thick, black!60] (0, -0.2) -- (0, 3.2) node[above, font=\small] {Trust Model};

%% --- Axis labels ---
\node[font=\tiny, below, text=black!50] at (0.5, -0.1) {Assistive};
\node[font=\tiny, below, text=black!50] at (1.5, -0.1) {Semi-Auton.};
\node[font=\tiny, below, text=black!50] at (2.7, -0.1) {Fully Auton.};

\node[font=\tiny, left, text=black!50, rotate=90, anchor=south] at (-0.1, 0.4) {Custodial};
\node[font=\tiny, left, text=black!50, rotate=90, anchor=south] at (-0.1, 1.1) {TEE};
\node[font=\tiny, left, text=black!50, rotate=90, anchor=south] at (-0.1, 1.8) {Crypto};
\node[font=\tiny, left, text=black!50, rotate=90, anchor=south] at (-0.1, 2.5) {Econ.\ Game};

%% --- Dense clusters (filled dots, A-Only = blue, B-Only = orange) ---
%% Cluster 1: (Assistive, Crypto, A-Only) — auditing tools
\node[dot, blue!60] at (0.4, 1.9) {};
\node[dot, blue!60] at (0.55, 1.75) {};
\node[dot, blue!60] at (0.5, 2.05) {};
\node[dot, blue!50] at (0.65, 1.95) {};
\node[lbl, blue!70, below right] at (0.7, 1.7) {Auditing\\tools};

%% Cluster 2: (Semi-Auton, Econ Game, A-Only) — DeFi agents
\node[dot, blue!60] at (1.4, 2.4) {};
\node[dot, blue!60] at (1.55, 2.55) {};
\node[dot, blue!60] at (1.6, 2.35) {};
\node[dot, blue!50] at (1.45, 2.6) {};
\node[dot, blue!50] at (1.7, 2.5) {};
\node[lbl, blue!70, right] at (1.75, 2.55) {DeFi\\agents};

%% Cluster 3: (Assistive, TEE, A-Only) — inference services
\node[dot, blue!60] at (0.4, 1.1) {};
\node[dot, blue!50] at (0.55, 1.0) {};
\node[dot, blue!50] at (0.5, 1.2) {};
\node[lbl, blue!70, right] at (0.65, 1.05) {Inference\\services};

%% Sparse B-Only dots (orange)
\node[dot, orange!60] at (0.5, 2.5) {};
\node[lbl, orange!70, above] at (0.5, 2.6) {Security\\monitors};

\node[dot, orange!60] at (1.0, 0.5) {};
\node[lbl, orange!70, right] at (1.05, 0.4) {DAO\\voters};

%% --- Opportunity regions (empty, dashed) ---
%% (Fully Auton, Crypto, Bidir)
\draw[region, red!50, fill=red!5] (2.3, 1.5) ellipse (0.5 and 0.45);
\node[lbl, red!60] at (2.3, 1.5) {\textbf{Gap 1}\\Verified\\bidir.\ agent};

%% (Semi-Auton, *, B-Only)
\draw[region, red!50, fill=red!5] (1.5, 0.7) ellipse (0.45 and 0.35);
\node[lbl, red!60] at (1.5, 0.7) {\textbf{Gap 2}\\AI\\validators};

%% (Fully Auton, Econ, Bidir)
\draw[region, red!50, fill=red!5] (2.7, 2.6) ellipse (0.4 and 0.35);
\node[lbl, red!60] at (2.7, 2.6) {\textbf{Gap 3}\\Self-sust.\\agents};

%% --- Legend ---
\node[dot, blue!60] at (0.1, 3.0) {};
\node[font=\tiny, right, blue!70] at (0.2, 3.0) {Dir.~A};
\node[dot, orange!60] at (1.0, 3.0) {};
\node[font=\tiny, right, orange!70] at (1.1, 3.0) {Dir.~B};
\draw[region, red!50] (1.95, 2.95) ellipse (0.15 and 0.1);
\node[font=\tiny, right, red!60] at (2.15, 2.95) {Opportunity gap};

% %% --- Cross-cutting concerns (outside the 2D space) ---
% \draw[densely dashed, thick, black!35, rounded corners=3pt]
%     (-0.2, -0.75) rectangle (3.4, -0.45);
% \node[font=\tiny\itshape, text=black!60] at (1.6, -0.6)
%     {Cross-cutting: Privacy, Cross-Chain, Legal (span all regions)};

\end{tikzpicture}
\caption{Taxonomy space (Autonomy $\times$ Trust Model). Blue/orange dots: B$\to$A / A$\to$B projects; cluster dot counts are illustrative, not proportional to actual project numbers. Red dashed ellipses (Gap~1--3) mark under-explored regions.}
\label{fig:taxonomy-space}
\end{figure}

These three gaps anticipate the research priorities formalized in \S\ref{sec:open-problems}: Gap~2 is taken up directly as OP4 (BFT for AI validators), Gap~1 spans OP1 (verifiable LLM inference) and OP6 (intent-to-proof pipelines), and Gap~3 maps to OP7 (token engineering for agent economics). The remaining open problems, namely permission verification (OP2), AI--human governance equilibria (OP3), inter-agent game theory (OP5), and legal frameworks for on-chain agents (OP8), are cross-cutting concerns that do not localize to any single region of the Autonomy$\times$Trust Model space, and are therefore shown only in \Cref{fig:gap-map} rather than as regions of \Cref{fig:taxonomy-space}.

The widest gap between industry activity and academic analysis falls in the (Semi-Autonomous, Economic Game, A-Only) region: dozens of DeFi trading agents exist in production, but virtually no formal analysis addresses their security properties, incentive compatibility, or failure modes. \Cref{fig:taxonomy-space} maps representative clusters and gaps in this three-dimensional space.

%% ===== Fig 7: Taxonomy Space (2D projection) =====

%% ==========================================================
%% SECTION 8: OPEN PROBLEMS
%% ==========================================================
\section{Open Problems and Research Agenda}\label{sec:open-problems}

\begin{figure}[t]
\centering
\begin{tikzpicture}[
    x=1.8cm, y=1.8cm,
    bubble/.style={circle, draw, thick, align=center, font=\tiny, inner sep=1pt},
    axlbl/.style={font=\small, text=black!70},
]

%% --- Axes ---
\draw[->, thick, black!50] (-0.2, 0) -- (4.2, 0) node[right, axlbl] {Difficulty};
\draw[->, thick, black!50] (0, -0.2) -- (0, 4.0) node[above, axlbl] {Impact};

%% --- Axis ticks ---
\node[font=\tiny, below, text=black!40] at (1, -0.1) {Low};
\node[font=\tiny, below, text=black!40] at (2.5, -0.1) {Medium};
\node[font=\tiny, below, text=black!40] at (3.8, -0.1) {High};
\node[font=\tiny, left, text=black!40] at (-0.1, 1) {Low};
\node[font=\tiny, left, text=black!40] at (-0.1, 2.3) {Medium};
\node[font=\tiny, left, text=black!40] at (-0.1, 3.5) {High};

%% --- Grid ---
\draw[black!8] (0,0) grid[step=0.5] (4,3.8);

%% --- Quadrant shading ---
\fill[green!4] (0,0) rectangle (2,2);
\fill[yellow!5] (2,0) rectangle (4,2);
\fill[yellow!5] (0,2) rectangle (2,3.8);
\fill[red!4] (2,2) rectangle (4,3.8);

%% --- OP bubbles ---
%% Color legend:
%%   Red shades    = Assurance bounds (OP1, OP4)
%%   Blue shades   = Protocol invariants (OP2, OP9)
%%   Orange shades = Mechanism design objectives (OP3, OP5, OP6, OP7, OP8)

%% OP1: Verifiable LLM inference — very high difficulty, very high impact
\node[bubble, fill=red!20, minimum size=0.85cm] (op1) at (3.5, 3.5) {OP1\\zkLLM};

%% OP2: Permission verification — medium-low difficulty, high impact
\node[bubble, fill=blue!18, minimum size=0.7cm] (op2) at (1.6, 3.3) {OP2\\Perm.};

%% OP3: Governance equilibria — high difficulty, high impact
\node[bubble, fill=orange!22, minimum size=0.75cm] (op3) at (3.0, 3.2) {OP3\\Gov.};

%% OP4: BFT for AI — high difficulty, medium-high impact
\node[bubble, fill=red!15, minimum size=0.7cm] (op4) at (3.4, 2.6) {OP4\\BFT};

%% OP5: Inter-agent game theory — medium-high difficulty, medium impact
\node[bubble, fill=orange!15, minimum size=0.65cm] (op5) at (2.7, 2.0) {OP5\\Game};

%% OP6: Intent-to-proof — high difficulty, medium impact
\node[bubble, fill=orange!15, minimum size=0.65cm] (op6) at (3.2, 1.7) {OP6\\Intent};

%% OP7: Token engineering — medium difficulty, medium impact
\node[bubble, fill=orange!18, minimum size=0.65cm] (op7) at (2.5, 2.35) {OP7\\Token};

%% OP8: Legal framework — low-medium difficulty, medium-high impact
\node[bubble, fill=orange!12, minimum size=0.65cm] (op8) at (1.7, 2.9) {OP8\\Legal};

%% OP9: Sybil-resistant agent identity — medium difficulty, high impact
\node[bubble, fill=blue!22, minimum size=0.75cm] (op9) at (2.25, 3.15) {OP9\\Sybil};

%% --- Quadrant annotations ---
\node[font=\tiny\itshape, text=black!30] at (1, 0.4) {Quick wins};
\node[font=\tiny\itshape, text=black!30] at (3.2, 0.4) {Hard but niche};
\node[font=\tiny\itshape, text=black!30, align=center] at (0.7, 3.5) {High-value,\\accessible};
\node[font=\tiny\itshape, text=black!30] at (3.6, 3.9) {Moonshots};

\end{tikzpicture}
\caption{Research gap map of the nine open problems by estimated difficulty and impact. Bubble color encodes \abim property type: \textcolor{red!60!black}{red} = assurance bounds, \textcolor{blue!60!black}{blue} = protocol invariants, \textcolor{orange!80!black}{orange} = mechanism design objectives.}
\label{fig:gap-map}
\end{figure}
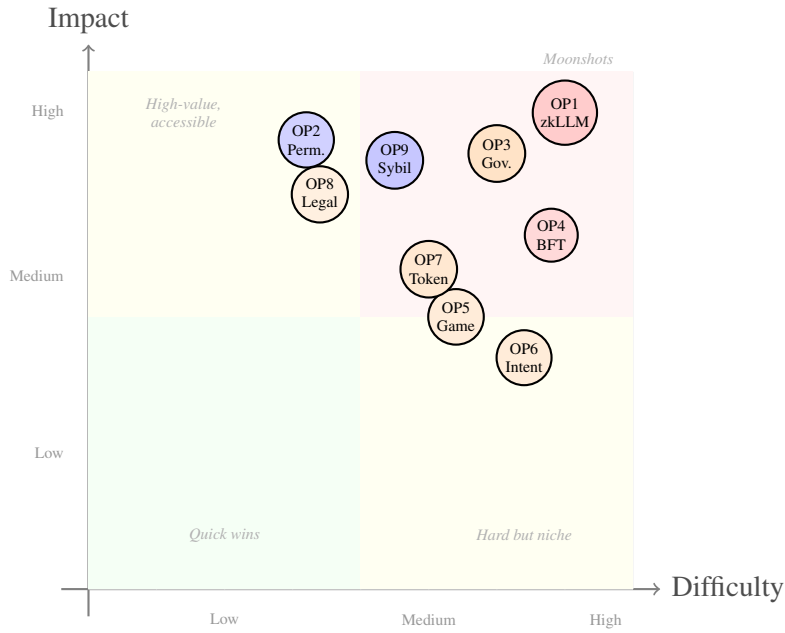

The analysis in the preceding sections identifies nine concrete open problems, each introduced in the context of the layer where it arises most naturally. Each problem corresponds to a specific \abim security property that current systems fail to satisfy (\Cref{tab:op-summary}). \Cref{fig:gap-map} visualizes the same information as a two-dimensional research gap map, with bubble color encoding the three \abim property types.

Three cross-cutting observations emerge from the aggregate view. First, the \emph{highest-impact problems} (OP1, OP3) involve bridging two different paradigms: cryptographic proof systems designed for deterministic computation must accommodate probabilistic AI models (OP1, targeting verification soundness \Cref{eq:tf}), and mechanism design theory must extend to heterogeneous human-AI populations (OP3, targeting bounded influence \Cref{eq:b3}). Second, the \emph{most accessible problems} (OP2, OP7, OP8, OP9) are those where existing theoretical tools, such as formal verification, token economics, legal theory, and proof-of-personhood, can be directly applied to the agent-blockchain setting with appropriate adaptation. Third, \emph{no single problem can be solved in isolation}, and the \abim property dependencies explain why: verification soundness (\Cref{eq:tf}) is a prerequisite for both validator verifiability (\Cref{eq:b2}, targeted by OP4) and solver faithfulness (\Cref{eq:a3}, targeted by OP6), while bounded influence (\Cref{eq:b3}, targeted by OP3 and OP8) depends on identity uniqueness (\Cref{eq:a1}, targeted by OP9). This dependency structure, formalized through \abim, should inform research prioritization: OP1 and OP9 are upstream bottlenecks whose resolution unblocks multiple downstream problems.

\begin{table}[t]
\caption{The nine open problems with their framework layer and targeted \abim property. Difficulty and impact are plotted in \Cref{fig:gap-map}.}\label{tab:op-summary}
\centering
\small
\begin{tabular}{@{}clll@{}}
\toprule
\textbf{\#} & \textbf{Layer} & \textbf{\abim Property} & \textbf{Problem} \\
\midrule
OP1 & TF & (\ref{eq:tf}) Verification Soundness & Practical verifiable LLM inference \\
OP2 & A2 & (\ref{eq:a2}) Enforcement Completeness & Formal verification of permission policies \\
OP3 & B3 & (\ref{eq:b3}) Bounded Influence & AI-human hybrid governance equilibria \\
OP4 & B2 & (\ref{eq:b2}) Validator Verifiability & BFT for AI validators \\
OP5 & A3 & (\ref{eq:a3}) Solver Faithfulness & Inter-agent game theory \\
OP6 & A3 & (\ref{eq:a3}) Solver Faithfulness & End-to-end intent-to-proof pipeline \\
OP7 & A4 & (\ref{eq:a4}) Incentive Compatibility & Token engineering for agent economics \\
OP8 & B3 & (\ref{eq:b3}) Bounded Influence & Legal framework for on-chain AI agents \\
OP9 & A1 & (\ref{eq:a1}) Identity Uniqueness & Sybil-resistant agent identity \\
\bottomrule
\end{tabular}
\end{table}

%% ===== Fig 8: Research Gap Map =====

%% ==========================================================
%% SECTION 9: CONCLUSION
%% ==========================================================
\section{Conclusion}\label{sec:conclusion}

This paper systematizes the rapidly growing intersection of autonomous AI agents and blockchain technology through a bidirectional trust framework. By cataloguing 70 EIPs/ERCs (App.~\ref{app:eip-catalog}), examining 20 representative industry projects, and reviewing 118 academic papers, and mapping them to our framework layers, we have identified both the substantial progress that has been made and the significant gaps that remain.

Several findings stand out. On the standards side, the agent-specific EIP ecosystem is overwhelmingly immature: only 2 of 13 direct AI/agent standards have reached Final, and no delegation standard has achieved Final status. On the academic side, the asymmetry between the two directions is stark: the \dirA direction (blockchain as infrastructure for agents) has attracted considerable industry attention and a growing body of standards, but the \dirB direction (agents as participants in consensus and governance) represents an almost entirely open academic frontier, one that will become increasingly important as AI agents grow more capable and autonomous.

The tools we have introduced serve complementary roles: the \abim formal model specifies what each layer \emph{must} guarantee (eight security properties, one per layer, classified as protocol invariants, assurance bounds, or mechanism design objectives), the five-dimensional evaluation framework measures how well current mechanisms \emph{do} satisfy those guarantees, and the three-dimensional taxonomy maps the ecosystem's coverage and blind spots. Together they are meant to serve as shared vocabulary for a research community that currently spans security, systems, cryptography, economics, and law, but often lacks a common frame of reference. The \abim compliance analysis (\Cref{tab:abim-compliance}) reveals that no property is fully satisfied by any deployed system, and the cross-layer dependencies among properties (notably \Cref{eq:b3} presupposing \Cref{eq:a1}) show that progress on governance bounds requires prior progress on identity infrastructure. The nine open problems we identify, each anchored to a specific \abim property, range from near-term engineering challenges (practical verifiable inference, targeting \Cref{eq:tf}; Sybil-resistant agent identity, targeting \Cref{eq:a1}) to long-term theoretical questions (governance equilibria, targeting \Cref{eq:b3}) and interdisciplinary frontiers (legal personhood for on-chain agents). We hope this systematization helps researchers navigate the design space and identify the problems most worth solving.

%% ==========================================================
%% REFERENCES
%% ==========================================================
% \bibliographystyle{ACM-Reference-Format}
% \bibliography{ref}
\bibliographystyle{aiasbst}
\bibliography{ref}

%% ==========================================================
%% APPENDIX (after references)
%% ==========================================================
\appendix
\section{Complete EIP/ERC Catalog}\label{app:eip-catalog}

\Cref{tab:eip-catalog-1,tab:eip-catalog-2} list all 70 EIPs and ERCs analyzed in this survey, organized into eight categories: \textbf{A}~Direct AI/Agent, \textbf{B}~Account Abstraction, \textbf{C}~Intent/Meta-Transaction, \textbf{D}~Delegation/Permission, \textbf{E}~Automation/Batch, \textbf{F}~Oracle/Off-chain, \textbf{G}~Smart Wallet, \textbf{H}~Identity/Signature. The ``Layer'' column maps each standard to the framework layer it primarily supports: TF~=~Trust Foundation, A1--A4~=~B\,$\to$\,A layers, B1~=~AI+Security, $\times$~=~cross-cutting.

\begin{table}[t]
\caption{Complete EIP/ERC catalog, Categories A--D (38 standards).}
\label{tab:eip-catalog-1}
\centering
% \tiny
\setlength{\tabcolsep}{2pt}
\begin{tabular*}{\columnwidth}{@{\extracolsep{\fill}}lllcc@{}}
\toprule
\textbf{ID} & \textbf{Title} & \textbf{Status} & \textbf{Category} & \textbf{Layer} \\
\midrule
\multicolumn{5}{@{}l}{\textit{A --- Direct AI/Agent (13)}} \\
ERC-7007 & Verifiable AI-Generated Content Token & Final & A & TF \\
ERC-7517 & Content Consent for AI/ML Data Mining & Draft & A & $\times$ \\
ERC-7641 & Intrinsic RevShare Token & Draft & A & A4 \\
ERC-7649 & Bonding Curve-Embedded Liquidity for NFTs & Draft & A & A4 \\
ERC-7662 & AI Agent NFTs & Draft & A & A4 \\
ERC-7845 & Universal Orchestrator RPC & Draft & A & A3 \\
ERC-7857 & AI Agents NFT with Private Metadata & Final & A & A1 \\
ERC-7992 & Verifiable ML Model Inference (ZKML) & Draft & A & TF \\
ERC-8004 & Trustless Agents & Draft & A & A1 \\
ERC-8033 & Agent Council Oracles & Draft & A & B1 \\
ERC-8121 & Cross-Chain Function Calls via Hooks & Draft & A & $\times$ \\
ERC-8126 & AI Agent Registration and Verification & Review & A & A1 \\
EIP-7911 & Scaling Ethereum with Perceptron Tree ZKP & Draft & A & TF \\
\midrule
\multicolumn{5}{@{}l}{\textit{B --- Account Abstraction (11)}} \\
EIP-86 & Abstraction of Transaction Origin and Signature & Stagnant & B & A1 \\
EIP-2938 & Account Abstraction & Withdrawn & B & A1 \\
ERC-4337 & Account Abstraction Using Alt Mempool & Review & B & A1 \\
ERC-5189 & Account Abstraction via Endorsed Operations & Draft & B & A1 \\
ERC-7562 & AA Validation Scope Rules & Draft & B & A1 \\
ERC-7679 & UserOperation Builder & Draft & B & A1 \\
EIP-7701 & Native Account Abstraction & Stagnant & B & A1 \\
EIP-7702 & Set Code for EOAs & Final & B & A1 \\
ERC-7766 & Signature Aggregation for ERC-4337 & Draft & B & A1 \\
ERC-7769 & JSON-RPC API for ERC-4337 & Draft & B & A1 \\
ERC-7902 & Wallet Capabilities for AA & Draft & B & A1 \\
\midrule
\multicolumn{5}{@{}l}{\textit{C --- Intent / Meta-Transaction (8)}} \\
ERC-1077 & Gas Relay for Contract Calls & Stagnant & C & A3 \\
ERC-1613 & Gas Stations Network & Stagnant & C & A3 \\
ERC-2771 & Secure Protocol for Native Meta Transactions & Final & C & A3 \\
ERC-3005 & Batched Meta Transactions & Stagnant & C & A3 \\
ERC-3009 & Transfer With Authorization & Draft & C & A3 \\
ERC-7521 & General Intents for Smart Contract Wallets & Draft & C & A3 \\
ERC-7683 & Cross Chain Intents & Draft & C & A3 \\
ERC-7806 & Minimal Intent-Centric EOA Smart Account & Draft & C & A3 \\
\midrule
\multicolumn{5}{@{}l}{\textit{D --- Delegation / Permission (6)}} \\
EIP-3074 & AUTH and AUTHCALL Opcodes & Withdrawn & D & A2 \\
ERC-5573 & Sign-In with Ethereum Capabilities, ReCaps & Draft & D & A2 \\
ERC-5639 & Delegation Registry & Review & D & A2 \\
ERC-7710 & Smart Contract Delegation & Draft & D & A2 \\
ERC-7715 & Request Permissions from Wallets & Draft & D & A2 \\
ERC-7741 & Authorize Operator & Draft & D & A2 \\
\bottomrule
\end{tabular*}
\end{table}

\clearpage

\begin{table}[!htbp]
\caption{Complete EIP/ERC catalog, Categories E--H (32 standards).}
\label{tab:eip-catalog-2}
\centering
% \tiny
\setlength{\tabcolsep}{2pt}
\begin{tabular*}{\columnwidth}{@{\extracolsep{\fill}}lllcc@{}}
\toprule
\textbf{ID} & \textbf{Title} & \textbf{Status} & \textbf{Category} & \textbf{Layer} \\
\midrule
\multicolumn{5}{@{}l}{\textit{E --- Automation / Batch Execution (5)}} \\
EIP-2803 & Rich Transactions & Stagnant & E & A3 \\
EIP-5792 & Wallet Call API & Final & E & A2 \\
EIP-5806 & Delegate Transaction & Stagnant & E & A3 \\
ERC-7821 & Minimal Batch Executor Interface & Draft & E & A2 \\
ERC-7836 & Wallet Call Preparation API & Draft & E & A2 \\
\midrule
\multicolumn{5}{@{}l}{\textit{F --- Oracle / Off-chain Computation (2)}} \\
EIP-7543 & EVM Arbitrary Precision Decimal Math & Stagnant & F & TF \\
ERC-7412 & On-Demand Off-Chain Data Retrieval & Draft & F & TF \\
\midrule
\multicolumn{5}{@{}l}{\textit{G --- Smart Wallet / Programmable Account (12)}} \\
EIP-5003 & Insert Code into EOAs with AUTHUSURP & Withdrawn & G & A1 \\
ERC-6551 & Non-fungible Token Bound Accounts & Review & G & A1 \\
ERC-6900 & Modular Smart Contract Accounts & Draft & G & A2 \\
ERC-7405 & Portable Smart Contract Accounts & Draft & G & A1 \\
ERC-7484 & Registry Extension for ERC-7579 & Draft & G & A2 \\
ERC-7579 & Minimal Modular Smart Accounts & Draft & G & A2 \\
ERC-7582 & Modular Accounts with Delegated Validation & Draft & G & A2 \\
ERC-7779 & Interoperable Delegated Accounts & Draft & G & A1 \\
ERC-7780 & Validation Module Extension for ERC-7579 & Draft & G & A2 \\
ERC-7895 & API for Hierarchical Accounts & Draft & G & A1 \\
ERC-7897 & Wallet-Linked Services for Smart Accounts & Withdrawn & G & A1 \\
ERC-7947 & Account Abstraction Recovery Interface & Draft & G & A1 \\
\midrule
\multicolumn{5}{@{}l}{\textit{H --- Identity / Verification / Signature (13)}} \\
ERC-1271 & Standard Signature Validation for Contracts & Final & H & A1 \\
ERC-2612 & Permit Extension for EIP-20 Signed Approvals & Final & H & A3 \\
ERC-4361 & Sign-In with Ethereum & Final & H & A1 \\
ERC-7522 & OIDC ZK Verifier for AA Account & Draft & H & A1 \\
ERC-7555 & Single Sign-On for Account Discovery & Draft & H & A1 \\
ERC-7597 & Signature Validation Extension for Permit & Draft & H & A2 \\
ERC-7677 & Paymaster Web Service Capability & Review & H & A3 \\
ERC-7739 & Readable Typed Signatures for Smart Accounts & Draft & H & A1 \\
ERC-7803 & EIP-712 Extensions for Account Abstraction & Draft & H & A1 \\
ERC-7812 & ZK Identity Registry & Review & H & A1 \\
ERC-7913 & Signature Verifiers & Final & H & A1 \\
ERC-7964 & Crosschain EIP-712 Signatures & Draft & H & $\times$ \\
ERC-7969 & DKIM Registry & Draft & H & A1 \\
\bottomrule
\end{tabular*}
\end{table}

\end{document}